# The exon junction complex undergoes a compositional switch that alters mRNP structure and nonsense-mediated mRNA decay activity

Justin W. Mabin[1,5,6], Lauren A. Woodward[1,5,6], Robert Patton[3,5,6], Zhongxia Yi[1,5], Mengxuan Jia[2], Vicki Wysocki[2,5], Ralf Bundschuh[3,4,5], Guramrit Singh[1,5,7*]

[1]Department of Molecular Genetics, The Ohio State University, Columbus, OH 43210
[2]Department of Chemistry and Biochemistry, The Ohio State University, Columbus, OH 43210
[3]Department of Physics, The Ohio State University, Columbus, OH 43210
[4]Division of Hematology, Department of Internal Medicine, The Ohio State University, Columbus, OH 43210
[5]Center for RNA Biology, The Ohio State University, Columbus, OH 43210
[6]These authors equally contributed to this work
[7]Lead contact

*Correspondence: Guramrit Singh (singh.734@osu.edu)

**SUMMARY**

The exon junction complex (EJC) deposited upstream of mRNA exon junctions shapes structure, composition and fate of spliced mRNA ribonucleoprotein particles (mRNPs). To achieve this, the EJC core nucleates assembly of a dynamic shell of peripheral proteins that function in diverse post-transcriptional processes. To illuminate consequences of EJC composition change, we purified EJCs from human cells via peripheral proteins RNPS1 and CASC3. We show that EJC originates as an SR-rich mega-dalton sized RNP that contains RNPS1 but lacks CASC3. After mRNP export to the cytoplasm and before translation, the EJC undergoes a remarkable compositional and structural remodeling into an SR-devoid monomeric complex that contains CASC3. Surprisingly, RNPS1 is important for nonsense-mediated mRNA decay (NMD) in general whereas CASC3 is needed for NMD of only select mRNAs. The promotion of switch to CASC3-EJC slows down NMD. Overall, the EJC compositional switch dramatically alters mRNP structure and specifies two distinct phases of EJC-dependent NMD.

**INTRODUCTION**

From the time of their birth during transcription until their eventual demise following translation, messenger RNAs (mRNA) exist decorated with proteins as mRNA-protein particles, or mRNPs (Gehring et al., 2017; Singh et al., 2015). The vast protein complement of mRNPs has been recently illuminated (Baltz et al., 2012; Castello et al., 2012; Hentze et al., 2018), and is believed to change as mRNPs progress through various life stages. However, the understanding of mechanisms and consequences of mRNP composition change remains confined to only a handful of its components. For example, mRNA export adapters are removed upon mRNP export to provide directionality to mRNP metabolic pathways, and the nuclear cap and poly(A)-tail binding proteins are exchanged for their cytoplasmic counterparts after mRNP export to promote translation (Singh et al., 2015, and references therein). When, where and how the multitude of mRNP components change during it's lifetime, and how such changes impact mRNP function remain largely unknown.

A key component of all spliced mRNPs is the exon junction complex (EJC), which assembles during pre-mRNA splicing 24 nucleotides (nt) upstream of exon-exon junctions (Boehm and Gehring, 2016; Le Hir et al., 2016; Woodward et al., 2017). Once deposited, the EJC enhances gene expression at several post-transcriptional steps including pre-mRNA splicing, mRNA export, mRNA transport and localization, and translation. If an EJC remains bound to an mRNA downstream of a ribosome terminating translation, it stimulates rapid degradation of the mRNA via nonsense-mediated mRNA decay (NMD). The stable EJC core forms when RNA bound EIF4A3 is locked in place by RBM8A (also known as Y14) and MAGOH. It has been proposed that this trimeric complex is joined by a fourth protein CASC3 (also known as MLN51 or Barenstz) to form a stable tetrameric core (Boehm and Gehring, 2016; Hauer et al., 2016; Le Hir et al., 2016; Tange et al., 2005). However, more recent evidence suggests that CASC3 may not be present in all EJCs and may not be necessary for all EJC functions (Mao et al., 2017; Singh et al., 2012). Nonetheless, the stable EJC core is bound by a dynamic shell of peripheral EJC proteins such as pre-mRNA splicing factors (e.g. SRm160, RNPS1), mRNA export proteins (e.g. the TREX complex), translation factors (e.g. SKAR) and NMD factors (e.g. UPF3B) (Boehm and Gehring, 2016; Le Hir et al., 2016; Woodward et al., 2017). Some peripheral EJC proteins share similar functions and yet may act on different mRNAs; e.g. RNPS1 and CASC3 can both enhance NMD but may have distinct mRNA targets (Gehring et al., 2005). Thus, the peripheral EJC shell may vary between mRNPs leading to compositionally distinct mRNPs, an idea that has largely remained untested.

We previously showed that within spliced mRNPs EJCs interact with one another as well as with several SR and SR-like proteins to assemble into mega-dalton sized RNPs (Singh et al., 2012). These stable mega RNPs ensheath RNA well beyond the canonical EJC deposition site leading to RNA footprints ranging from 150-200 nt in length, suggesting that the RNA polymer within these complexes is packaged resulting in overall compact mRNP structure. Such a compact structure may facilitate mRNP navigation of the intranuclear environment, export through the nuclear pore and transport within the cytoplasm to arrive at its site of translation. Eventually, the mRNA within mRNPs must be unpacked to allow access to the translation machinery. How long do mRNPs exist in their compact states, and when, where and how are they unfurled remains yet to be understood.

Our previous observation that in human embryonic kidney (HEK293) cells CASC3 and many peripheral EJC factors are substoichiometric to the EJC core (Singh et al., 2012) spurred us to investigate variability in EJC composition. Here we use EJC purification via substoichiometric factors to reveal that EJCs first assemble into SR-rich mega-dalton sized RNPs, and then undergo a compositional switch into SR-devoid monomeric CASC3-containing EJC. The translationally repressed mRNPs, particularly those encoding ribosomal proteins, accumulate with the CASC3-bound form of the EJC. Although both the EJC forms remain active in NMD, RNPS1, a component of the SR-rich EJCs, is crucial for all tested NMD events whereas CASC3, a constituent of SR-devoid EJC, is dispensable for NMD of many transcripts. Our findings reveal a new step in the mRNP life cycle wherein EJCs, and by extension mRNPs, undergo a remarkable compositional switch that alters the mRNP structure and specifies two distinct phases of EJC dependent NMD.

## RESULTS

**RNPS1 and CASC3 associate with the EJC core in a mutually exclusive manner**

We reasoned that substoichiometric EJC proteins may not interact with all EJC cores, and therefore, some of them may not interact with each other leading to variable EJC composition. To test such a prediction, we carried out immunoprecipitations (IP) of either endogenous core factor EIF4A3 or the substoichiometric EJC proteins RNPS1 and CASC3 from RNase A-treated HEK293 total cell extracts. As expected, EIF4A3 IP enriches EJC core as well as all peripheral proteins tested (Figure 1A, lane 3). In contrast, the IPs of substoichiometric factors enriched only distinct sets of proteins. CASC3 immunopurified the EJC core proteins (EIF4A3, RBM8A and MAGOH) but not the peripheral proteins ACIN1 and SAP18 (Figure 1A, lane 4; RNPS1 could not be blotted for as it co-migrates with IgG heavy chain). Conversely, following RNPS1 IP, the EJC core proteins as well as its known binding partner SAP18 (Murachelli et al., 2012; Tange et al., 2005) were enriched but CASC3 was undetected in these IPs. We saw an identical lack of co-IP between RNPS1 and CASC3 from cells that were crosslinked with formaldehyde prior to lysis (data not shown; also see below), suggesting that the lack of interaction between these substoichiometric factors does not result from their dissociation following cell lysis. We also carried out similar IPs from RNase-treated total extracts from mouse brain cortical slices, mouse embryonal carcinoma (P19) cells, and HeLa cells. Again, both RNPS1 and CASC3 efficiently co-IPed with EIF4A3 from these lysates but no co-IP was detected between RNPS1 and CASC3 (compare lanes 4 and 5 with lane 3 in Figures 1B, S1A and S1B). Therefore, we conclude that in mammalian cells RNPS1 and CASC3 exist in complex with the EJC core in a mutually exclusive manner, and term these compositionally distinct EJCs as alternate EJCs.

**Proteomic Analysis of Alternate EJCs**

To gain insights into alternate compositions of the EJC, we further characterized RNPS1 and CASC3-containing complexes using a bottom-up proteomics approach. We generated HEK293 cell lines stably expressing FLAG-tagged RNPS1 and CASC3 from a tetracycline inducible promoter where the proteins are induced at near endogenous levels (Figure S1C). FLAG IPs from RNase-treated total extracts of FLAG-RNPS1 and FLAG-CASC3 expressing cells confirmed that the tagged proteins also exist in mutually exclusive EJCs (Figure S1D). Importantly, proteomic analysis of FLAG-affinity purified alternate EJCs show almost complete lack of spectra corresponding to RNPS1 and CASC3 in the immunoprecipitates of the other alternate EJC factor (Figures 1C and S1E). In comparison, a FLAG-MAGOH IP enriches both CASC3 and RNPS1 as expected (Figures 1C, S1D and S1E).

We next compared the full complement of proteins associated with FLAG-RNPS1, FLAG-CASC3 and FLAG-MAGOH focusing on proteins >2-fold enriched in any one of the three EJC IPs as compared to the FLAG-only negative control (Table S1; see Experimental Procedures). Of the 59 proteins enriched in the two FLAG-RNPS1 biological replicates (Figure S2A), 38 are common with the 45 proteins identified in the two FLAG-MAGOH replicates (Figures S2A and S2B). Such a high degree of overlap suggests that RNPS1 containing complexes are compositionally similar to those purified via the EJC core. In comparison, among the two FLAG-CASC3 replicates the only common proteins are EIF4A3 and MAGOH (Figure S2A), which are also shared with FLAG-MAGOH (Figure S2C). Known EJC interactors such as UPF3B and PYM1 are identified in only one of the two FLAG-CASC3 replicates indicating less stable CASC3 association with these EJC factors.

A direct comparison of FLAG-RNPS1 and FLAG-CASC3 proteomes highlights their stark differences beyond the common EJC core (Figures 2A and 2B). Several SR and SR-like proteins are enriched in the RNPS1 and MAGOH proteomes but are completely absent from the CASC3 samples. Among the SR protein family, SRSF1, 6, 7 and 10 are reproducibly enriched in all RNPS1 and MAGOH samples, while all other canonical SR proteins, with the exception of non-shuttling SRSF2, are detected in at least one of the two replicates (Figure 2A, 2B, 2C and Table S1). Several SR-like proteins such as ACIN1, PNN, SRRM1 and SRRM2 are also highly enriched in MAGOH and RNPS1 IPs but are absent in CASC3 IPs (Figure 2 and Table S1). A weak but erratic signal for SRSF1 and SAP18 is seen in FLAG-CASC3 IPs (e.g. spectra for two SRSF1-derived peptides are observed in one of the FLAG-CASC3 samples; Figure 2B). However, their enrichment in CASC3-EJC is much weaker as compared to MAGOH or RNPS1 samples. Overall, these findings suggest an extensive interaction network between SR and SR-like proteins and the RNPS1-EJC but not the CASC3-EJC.

The proteomes of the three EJC factors can be readily classified based on protein functions in RNA biogenesis, their assembly/function within RNPs, or their protein sequence composition (Figure 2B). Many RNA biogenesis factors are specifically associated with RNPS1 (and MAGOH) but not with CASC3. These include transcription machinery components (e.g. RPB1, RPB2), transcriptional regulators (e.g. CD11A, CDK12) as well as RNA processing factors (e.g. NCBP1, FIP1). Consistent with EJC core assembly during pre-mRNA splicing, MAGOH and RNPS1 interactors also include U2, U4 and U6 snRNP components and nineteen complex subunits. None of the splicing components are seen to interact with CASC3. Considering that SR proteins also assemble onto nascent RNAs co-incident with transcription and splicing (Zhong et al., 2009), the EJC-RNPS1-SR interaction network likely originates during co-transcriptional mRNP biogenesis. In contrast, CASC3 most likely engages with the EJC only post-splicing, as previously suggested (Gehring et al., 2009a).

**Alternate EJCs differ in their higher order structure**

The enrichment of SR and SR-like proteins exclusively with RNPS1 suggested that RNPS1 EJCs are likely to resemble the previously described higher order EJCs (Singh et al., 2012). Indeed, glycerol gradient fractionation of RNase-treated complexes immunopurified via FLAG-RNPS1 shows that, like EJCs purified via FLAG-MAGOH, FLAG-RNPS1 EJCs contain both lower and higher molecular weight complexes (Figure 2D). On the other hand CASC3 is mainly detected in lower molecular weight complexes purified via FLAG-MAGOH (Figure 2D, compare CASC3 signal in fractions 2-10 and 22-24). Furthermore, complexes purified via FLAG-CASC3 are exclusively comprised of lower molecular weight complexes, likely to be EJC monomers (Figure 2D). These findings are consistent with the proteomic results and suggest that compositional distinctions between the two alternate EJCs give rise to two structurally distinct complexes.

**RNPS1 and CASC3 bind RNA via the EJC core with key distinctions**

We next identified the RNA binding sites for the two alternate EJC factors using RNA:protein immunoprecipitation in tandem (RIPiT) combined with high-throughput sequencing, or RIPiT-Seq (Singh et al., 2012, 2014). RIPiT-Seq entails tandem purification of two subunits of an RNP, and is well-suited to study EJC composition via sequential IP of its constant (e.g. EIF4A3, MAGOH) and variable (e.g. RNPS1, CASC3) components (Figure 3A). We carried out RIPiTs from HEK293 cells by either pulling first on FLAG-tagged alternate EJC factor followed by IP of an endogenous core factor, or vice versa. As a control, we also performed FLAG-MAGOH:EIF4A3 RIPiT-Seq. All *in vivo* EJC binding studies thus far have employed a short incubation with translation elongation inhibitor cycloheximide prior to cell lysis to limit EJC disassembly by translating ribosomes (Hauer et al., 2016; Saulière et al., 2012; Singh et al., 2012). However, to capture unperturbed, steady-state populations of RNPS1 and CASC3 bound EJCs (as in Figures 1 and 2), we performed RIPiT-Seq without translation inhibition.

As expected, RIPiTs for each of the two alternate EJCs specifically purified the targeted complex along with the EJC core factors and yielded abundant RNA footprints (Figure S3A). Strand-specific RIPiT-Seq libraries from ~35-60 nucleotide footprints yielded ~5.7 to 47 million reads, of which >80% mapped uniquely to the human genome (hg38; Table S2). Genic read counts are highly correlated between RIPiTs where the order of IP of EJC core and alternate factors was reversed (Figures S3B and S3C). We found that RNPS1-EJC core interaction is susceptible to NaCl concentrations greater than 250 mM whereas CASC3-EJC core interaction persists up to 600 mM NaCl (data not shown). Therefore, to preserve labile interactions we performed alternate EJC RIPiTs from cells cross-linked with formaldehyde before cell lysis (Figure S3D). We observed a strong correlation for genic read counts between crosslinked and uncrosslinked samples (Figures S3E and S3F). The analysis presented below is from two well-correlated biological replicates of formaldehyde crosslinked RIPiT-Seq datasets of RNPS1- and CASC3-EJC (Figures S3G and S3H).

Consistent with EJC deposition on spliced RNAs, alternate EJC footprints are enriched in exonic sequences from multi-exon genes (Figure S3I). Along individual exons, footprints of both alternate EJCs mainly occur close to exon 3' ends (Figure 3B). Indeed, a meta-exon analysis shows that the alternate EJC factors bind mainly to the canonical EJC binding site 24 nt upstream of exon junctions (Figure 3C) (Hauer et al., 2016; Saulière et al., 2012; Singh et al., 2012). Notably, the alternate EJC footprint enrichment at the canonical EJC site is irrespective of whether the alternate EJC factors are IPed first or second during the RIPiT procedure (Figure S3J). Thus, both RNPS1 and CASC3 mainly bind to RNA via EJCs at the canonical site. The location of the 5'- and 3'-ends of the alternate EJC footprint reads shows that both alternate EJCs have a similar footprint on RNA as each blocks positions -26 to -19 nt from RNase I cleavage (Figure 3D). Our findings are consistent with previous work that showed that CASC3 mainly binds to the canonical EJC site (Hauer et al., 2016). However, in contrast to this study, RNPS1 binding to canonical EJC sites is readily apparent (Figures 3B, 3C and 3D), which could reflect differences in the UV-crosslinking dependent CLIP-Seq approach used by Hauer *et al.* and the photo-crosslinking independent RIPiT-Seq. Altogether, our results suggest that both RNPS1 and CASC3 exist in complex with EJC at its canonical site.

While read densities for both alternate EJC factors are highest at the canonical EJC site, 47-62% of reads map outside of the canonical EJC site similar to previous estimates (Saulière et al., 2012; Singh et al., 2012). Non-canonical footprints are somewhat more prevalent in RNPS1-EJC as compared to CASC3-EJC (Figure S3K). As RNPS1-EJC is intimately associated with SR and SR-like proteins (Figure 2A and B), a k-mer enrichment analysis revealed a modest but significant enrichment of GA-rich 6-mers in RNPS1 over CASC3 footprints (Figures 3E). Such purine-rich sequences occur in binding sites of several SR proteins (SRSF1, SRSF4, Tra2a and b) (Änkö et al., 2012; Pandit et al., 2013; Tacke et al., 1998). There is also a small but significant enrichment of sequence motifs recognized by SRSF1 and SRSF9 in RNPS1 footprints as compared to CASC3 footprints or RNA-Seq reads (Figures 3F, 3G, S3L and S3M). We conclude that within spliced RNPs, RNPS1-containing EJC is engaged with SR and SR-like proteins, and other RNA binding proteins, which leads to co-enrichment of RNA binding sites of these proteins during RIPiT via RNPS1. Intriguingly, CG-rich 3-mers are the highest enriched k-mers in footprints of both alternate EJCs as compared to RNA-Seq (data not shown). The CG-rich 6-mers are also somewhat more enriched in RNPS1-EJC (Figure 3E). A CG-rich sequence was reported as an *in vitro* RBM8A binding motif indicating a yet to be determined relationship between the EJC core and CG-rich sequences (Ray et al., 2013).

**RNPS1 and CASC3 RNA occupancy changes with mRNP subcellular location**

Surprisingly, despite the mutually exclusive association of RNPS1 and CASC3 with the EJC core (Figures 1 and 2), the two proteins are often detected on the same sites on RNA leading to their similar apparent occupancy on individual exons as well as entire transcripts (Figures 3B, S4A and S4B). These results suggest that the two alternate EJC factors bind to two distinct pools of the same RNAs. To further reveal RNA binding patterns of the alternate EJC factors, we first identified exons differentially enriched in one or the other factor. Remarkably, if two or more exons of the same gene are differentially enriched in RNPS1 or CASC3 footprints, these exons are almost always enriched in the same alternate EJC factor (Figure 4A, $p < 0.00001$). This extremely tight linkage between EJC compositions of different exons of the same gene suggests that RNPS1 and CASC3 binding to EJC is determined at the level of the entire transcript and not of the individual exon. At any given time a transcript with multiple EJCs is likely to be homogeneously associated with either one or the other alternate factor. In support of this scenario, very weak RNA-dependent interaction is detected between the two alternate EJC factors (Figure S4C).

At steady state, RNPS1 is mainly localized to the nucleus whereas CASC3 is predominantly cytoplasmic although both proteins shuttle between the two compartments (Daguenet et al., 2012; Degot et al., 2002; Lykke-Andersen et al., 2001). We reasoned that different concentration of alternate EJC factors in the two subcellular compartments may mirror their EJC association and RNA occupancy. To test this possibility, we identified subsets of transcripts that are preferentially enriched in RNPS1- (242 transcripts) or CASC3-EJC (625 transcripts; Figure 4B), and compared their nuclear:cytoplasmic ratios based on estimates of subcellular RNA distribution in HEK293 cells (Neve et al., 2016). Indeed, the transcripts enriched in CASC3-EJC strongly localize to the cytoplasm (median % localized in cytoplasm = 76), whereas those preferentially bound to RNPS1 show more even distribution with a slight bias for nuclear localization (median % localized in cytoplasm = 48; Figures 4C and S4D). Thus, subcellular localization of spliced RNAs and EJC factors are important determinants of EJC composition.

We surmised that kinetics of mRNA maturation and in turn nuclear export will directly impact EJC composition. To test this idea, we compared alternate EJC occupancy of a group of genes whose nuclear poly(A)-tailed transcripts contain introns that undergo splicing much more slowly as compared to other introns in the same transcript (Boutz et al., 2015). Transcripts containing these detained introns (DI) are restricted to the nucleus in a mostly pre-processed state until DIs are spliced, and are therefore expected to have slower mRNA export rates. Of the genes that Boutz *et al.* found to contain DI in four human cell lines, 542 also contained DIs in HEK293 cells whereas a set of 389 transcripts do not contain DI in any human cell lines including HEK293 cells (data not shown). We find that DI-containing transcripts are significantly more enriched in RNPS1-EJC ($p=8.3 \times 10^{-13}$, Figures 4D and 4E) while DI lacking transcripts show an enrichment in CASC3-EJC ($p=1.8 \times 10^{-7}$). These results support our conclusion that EJC (and mRNP) compositional switch mainly occurs as they mature and enter the cytoplasm, and that the rate of mRNA progression to the cytoplasm alters the switch.

Despite a strong CASC3 enrichment on cytoplasmic RNAs, a quarter of all preferentially CASC3-bound RNAs are more nuclear (Figure 4C). Consistent with CASC3 shuttling into the nucleus (Daguenet et al., 2012; Degot et al., 2002), its footprints are abundantly detected on *XIST* RNA and several other spliced non-coding RNAs restricted or enriched in the nucleus (Figures 4F and S4D). Further, in HEK293 nuclear as well as cytoplasmic extracts both CASC3 and RNPS1 are found to assemble into EJCs (Figures S4E and S4F). These data suggest that while the bulk of RNPs may undergo the switch in EJC composition in the cytoplasm, some RNPs can undergo this change in the nucleus itself, possibly as a function of RNP half-life or due to a more active nuclear role of CASC3 in RNP biogenesis and function.

**Translation and mRNA decay kinetics impacts alternate EJC bound mRNA pools**

As EJCs are disassembled during translation (Dostie and Dreyfuss, 2002; Gehring et al., 2009b), abundant RNPS1 and CASC3 footprints at EJC deposition sites suggest that the bulk of mRNPs undergo the compositional switch before translation. To test how translation impacts occupancy of both alternate EJCs, and if their occupancy is influenced by rate at which mRNAs enter translation pool, we obtained RNPS1- and CASC3-EJC footprints from cells treated with cycloheximide (CHX), and compared them to alternate EJC footprints from untreated cells. When mRNAs bound to each alternate EJC are compared across the two conditions, CASC3 RNA occupancy shows a dramatic change (Figure 5A). In contrast, while changes in RNPS1-bound RNAs across two conditions shows the same trend as CASC3 ($R^2$=0.27, Figure 5A), the change is much less dramatic with RNPS1 occupancy changing significantly only on a handful of transcripts. These observations further support our conclusion that the RNPS1-bound form of the EJC precedes the CASC3-bound form. They also indicate that translation inhibition does not interfere with the compositional switch from RNPS1 to CASC3 but leads to accumulation of CASC3-bound mRNPs.

It is expected that poorly translated or translationally repressed mRNAs will be enriched in CASC3-EJC under normal conditions, and more efficiently translated mRNAs will be differentially enriched upon translation inhibition. To test this hypothesis, we obtained mRNA translation efficiency (TE) estimates based on ribosome footprint counts for each transcript normalized to its abundance from a human colorectal cancer cell line (Kiss et al., 2017). A comparison of the CASC3-EJC enriched transcripts from the two conditions however showed only a minor difference in their median TE (Figure S5A). A search for functionally related genes in the two sets revealed that each contains diverse groups (Figure 5B). Among the transcripts differentially bound to CASC3-EJC under normal conditions, the largest and most significantly enriched group encodes signal-peptide bearing secretory/membrane bound proteins (Figure 5B), which has a significantly higher TE as compared to all transcripts (Figure S5B). We reason that despite the higher TE of the "secrotome" (all secreted/membrane proteins as classified in (Jan et al., 2014)), transcripts with signal peptide may be enriched in CASC3-EJC because binding of the hydrophobic signal peptide to the signal recognition particle halts translation, which resumes only when the ribosome engages with the endoplasmic reticulum (ER) (Walter and Blobel, 1981). Presumably, the time before translation resumption on the ER allows capture of mRNPs where EJC composition has switched but the complex has not yet been disassembled by translation. When we excluded the secretome and considered only those transcripts that are translated in the cytosol (as defined in (Chen et al., 2011)), a comparison of TE of CASC3-EJC enriched transcripts under translation conducive and inhibitory conditions confirmed our initial hypothesis. As seen in Figure 5C, transcripts bound to CASC3-EJC in the absence of CHX have significantly lower median TE (-2.56) as compared to median TE of transcripts bound to CASC3-EJC in the presence of CHX (-2.08, $p=2.7 \times 10^{-4}$).

Another functional group enriched in CASC3-EJC under normal conditions comprises the ribosomal protein (RP)-coding mRNAs (Figure 5B). Strikingly, transcripts encoding ~50% of all cytosolic ribosomal subunit proteins as well as 13 mitochondrial ribosome subunits are among this set (Figure 5A). Although RP mRNAs are among the most highly translated in the cell, a sizeable fraction of RP mRNAs are known to exist in a dormant untranslated state that can enter the translation pool upon demand (Geyer et al., 1982; Meyuhas and Kahan, 2015; Patursky-Polischuk et al., 2009). Consistently, RP mRNAs have significantly low TE in human and mouse cells (Figures S5B and S5C). Furthermore, when transcripts differentially bound to RNPS1- versus CASC3-EJC are directly compared in normally translating cells, RP mRNAs are specifically enriched among CASC3-EJC bound transcripts (Figure 5D). Therefore, RP mRNPs, and perhaps other translationally repressed mRNPs, that linger in the untranslated state in the cytoplasm switch to and persist in the CASC3-bound form of the EJC. Such a possibility is further supported by significantly lower TE of CASC3-EJC enriched RNAs as compared to RNPS1-EJC bound transcripts under normal conditions (Figure 5E). Further, when mRNPs are forced to persist in an untranslated state upon CHX treatment, cytosol translated mRNAs show increased CASC3 occupancy whereas their RNPS1 occupancy is not affected (Figure S5D).

As previously reported (Hauer et al., 2016), RP mRNAs are depleted of CASC3-EJC upon translation inhibition (Figures 5A and 5D). Upon cycloheximide treatment, CASC3 occupancy is significantly reduced at canonical EJC sites of RP mRNAs as compared to non-RP mRNAs, which show an increase in CASC3 occupancy (Figure 5F, $p=2.5 \times 10^{-43}$). In comparison, RNPS1 occupancy on all transcripts modestly increases upon translation inhibition (Figure 5F, p=0.059). A paradoxical possibility is that the untranslated reserves of RP mRNAs may undergo translation when the cellular pool of free ribosomes is dramatically reduced upon CHX-mediated arrest of translating ribosomes on mRNAs. Intriguingly, a recent study in fission yeast found a similar contradictory increase in ribosome footprint densities on RP mRNAs upon cycloheximide treatment (Duncan and Mata, 2017).

The alternate EJC occupancy landscape is also impacted by mRNA decay kinetics. CASC3-EJC enriched RNAs have longer half-lives as compared to RNPS1-EJC enriched transcripts under both translation conducive (median $t_{1/2}$=4.6 hr vs. 5.9 hr, $p=3.1 \times 10^{-3}$) and inhibitory conditions (median $t_{1/2}$=3.4 hr vs. 4.8 hr, $p=5.3 \times 10^{-5}$, Figure 5G). Notably, RNAs enriched in both alternate EJCs upon CHX treatment have lower half-lives as compared to the corresponding cohorts enriched from normal conditions. Thus, EJC detection is enhanced on transcripts that are stabilized after CHX treatment. Such a conclusion is further supported by enrichment of functionally related groups of genes known to encode unstable transcripts (e.g. cell cycle, mRNA processing and DNA damage, (Schwanhäusser et al., 2011)) in CASC3-EJC upon translation inhibition (Figure 5B).

**RNPS1 is required for efficient NMD of all transcripts whereas CASC3 is dispensable for many**

Our data suggests that mRNPs arrive in the cytoplasm with RNPS1 and SR protein-containing EJCs, which are remodeled into CACS3-containing EJCs. We wanted to determine if alternate EJCs were equally important for NMD and/or if they have distinct targets as previously suggested (Gehring et al., 2005). In HEK293 cells with ~80% of RNPS1 mRNA and proteins depleted, a majority of endogenous NMD targets tested are significantly upregulated (Figures 6A and S6A). Surprisingly, however, in cells with ~85% of CASC3 depleted, NMD targets are largely unaffected; some others are only modestly upregulated, and only one out of thirteen is significantly upregulated. Notably, the RNAs that were upregulated upon CASC3 depletion are even more upregulated upon RNPS1 depletion. Most tested RNAs are significantly upregulated upon depletion of EIF4A3 and the central NMD factor UPF1 (Figure 6A and data not shown). These data suggest that RNPS1 function is required for efficient NMD of most endogenous NMD substrates whereas CASC3 may be needed for NMD of only select RNAs.

It is possible that the residual amount of CASC3 after its siRNA mediated depletion is sufficient to support NMD. To test this further, we depleted endogenous EIF4A3 protein to ~50% of its normal levels and supplemented cells with either a WT EIF4A3 (WT) or an EIF4A3 mutant (YRAA: Y205A, R206A) with much reduced CASC3 binding (Figures 6B, S6B, S6C, S6D, (Andersen et al., 2006; Ballut et al., 2005; Bono et al., 2006). The EIF4A3 knockdown results in >2-fold upregulation of all NMD targets tested (Figure 6B). We find that complementation of EIF4A3 knockdown cells with exogenous FLAG-EIF4A3 or FLAG-EIF4A3 YRAA proteins leads to nearly identical effect on NMD restoration (Figure 6B). Based on the degree of rescue, the NMD targets tested can be divided into two groups. The first group (to the left of vertical red dotted line in Figure 6B) shows a strong rescue of NMD upon complementation with both the wild-type or the mutant EIF4A3 proteins. Therefore, this group of transcripts can undergo NMD largely independent of CASC3. In the second group (to the right of vertical red dotted line in Figure 6B), neither the wild-type nor the mutant proteins can rescue NMD defect caused by the depletion of endogenous EIF4A3 protein. We noted that, unlike the first group (with exception of C1orf37), the second group is somewhat more sensitive to depletion of CASC3 (Figure 6A). This subset of mRNAs may have a more complex dependence on EIF4A3 levels.

In contrast to most endogenous NMD substrates, we found that knockdown of both RNPS1 and CASC3 in HeLa cells leads to accumulation of a well-known β-globin mRNA with a premature stop codon at codon 39 (β39, Figures 6C and S6E). Consistently, recent genome-wide screens have identified CASC3 as an effector of NMD of exogenous reporters (Alexandrov et al., 2017; Baird et al., 2018). Overall, our results show that, in human cells, several NMD targets can undergo CASC3 independent NMD whereas almost all tested transcripts depend on RNPS1 for their efficient NMD. On the other hand, exogenously (over) expressed NMD substrates may depend on both sequential EJC compositions for their efficient NMD.

**Increased CASC3 levels slow down NMD**

We next tested if overexpression of RNPS1 or CASC3 can tilt EJC composition toward one of the two alternate EJCs, and if such a change can impact NMD. CASC3 overexpression in HEK293 cells leads to several fold increase in CASC3 co-IP with EIF4A3 (Figures 7A, 7B) and RBM8A (Figures S7A and S7B), although no concomitant decrease is seen in RNPS1 co-precipitation with EJC core proteins. Consistent with this, manifold overexpression of RNPS1 did not cause any detectable change in levels of the two alternate factors in the EJC core IPs (Figures 7A, 7B and S7A and S7B). These results rule out a simple, direct competition between the two proteins for EJC core interaction. We next tested if increased CASC3 association with EJCs under elevated CASC3 levels affects NMD. Surprisingly, more than half of tested endogenous NMD targets are significantly upregulated upon CASC3 overexpression (Figure 7C). Similarly, the β39 mRNA exogenously expressed in HeLa cells is also modestly stabilized upon CASC3 overexpression (Figure 7D and S7C). As previously reported (Viegas et al., 2007), RNPS1 overexpression further downregulates β39 mRNA in HeLa cells (Figures S7C and S7D), indicating that NMD of this RNA occurs more efficiently when it is associated with early acting SR-rich EJC.

## DISCUSSION

The EJC is a cornerstone of all spliced mRNPs, and interacts with upwards of 50 proteins to connect the bound RNA to a wide variety of post-transcriptional events. The EJC is thus widely presumed to be "dynamic". By purifying EJC via key peripheral proteins, we demonstrate that a remarkable binary switch occurs in EJC's complement of bound proteins. Such an EJC composition change has important implications for mRNP structure and function including mRNA regulation via NMD (Figure 7E).

### EJC composition and mRNP structure

Our findings suggests that when EJCs first assemble during co-transcriptional splicing, the core complex consisting of EIF4A3, RBM8A and MAGOH engages with SR proteins and SR-like factors including RNPS1 (Figure 2). Within these complexes, RNPS1 is likely bound to both the EJC core as well as to the SR and SR-like proteins bound to their cognate binding sites on the RNA (Figure 3). This network of interactions bridges adjacent and distant stretches of mRNA, winding the mRNA up into a higher-order structure, which is characteristic of RNPs purified from human cells via the EJC core proteins or RNPS1 (Figure 2, (Singh et al., 2012)). Such higher-order mRNA packaging can create a compact RNP particle for its efficient intracellular transport. It is possible that assembly of these higher-order structures occur via multiple weak interactions among low complexity sequences (LCS) within EJC bound SR and SR-like proteins (Haynes and Iakoucheva, 2006; Kwon et al., 2014). It also remains to be tested if RNPS1, which possesses an SR-rich LCS, acts as a bridge between the EJC core and more distantly bound SR proteins. Our data also indicates that on a majority of mRNPs, these SR-rich and RNPS1-containing higher-order EJCs persist throughout their nuclear lifetime (Figure 4). When mRNPs arrive in the cytoplasm, the SR and SR-like proteins are evicted from all EJCs of an mRNP and the EJC is joined by CASC3 (Figure 4). It remains to be seen if CASC3 incorporation into an EJC causes its remodeling. Alternatively, active process(es) such as RNP modification via SR protein phosphorylation by cytoplasmic SR protein kinases (Zhou and Fu, 2013) or RNP remodeling by ATPases may precede CASC3 binding to EJC (Lee and Lykke-Andersen, 2013). What is clear is that CASC3-bound EJCs lose their higher order structure and exist as monomeric complexes at the sites where EJC cores were co-transcriptionally deposited. Thus, the switch in EJC composition from RNPS1 and SR-rich complexes to CASC3-bound complexes leads to a striking alteration in higher-order EJC, and possibly, mRNP structure (Figure 7E). The CASC3 bound form of the EJC following compositional switch is the main target of translation dependent disassembly although RNPS1-EJC may undergo some translation-dependent disassembly (Figure 5).

**CASC3-EJC and pre-translation mRNPs**

Our findings support the emerging view that CASC3 is not an obligate component of all EJC cores. A population of assembled EJCs, especially those early in their lifetime, completely lack CASC3 (Figures 1 and 2). Such a view of partial CASC3 dispensability for EJC structure and function is in agreement with findings from *Drosophila* where the assembled trimeric EJC core as well as RNPS1 and its partner ACIN1 are required for splicing of long or sub-optimal introns whereas CASC3 is not (Hayashi et al., 2014; Malone et al., 2014; Roignant and Treisman, 2010). A recent report that lack of CASC3 during mouse embryonic brain development results in phenotypes distinct from those caused by similar mutations of the other three core proteins also supports non-overlapping functions of CASC3 and the other core factors (Mao et al., 2017). We note that in the recently reported human spliceosome C* structure, CASC3 is bound to the trimeric core (Zhang et al., 2017). As the spliceosomes described in these structural studies were assembled *in vitro* in nuclear extracts, it is possible that CASC3 present in extracts can enter pre-assembled spliceosomes and interact with EJC. Consistently, in the human spliceosome C* structure one of the two CASC3 binding surfaces on EIF4A3 is exposed and available for CASC3 interaction. Still, a possibility remains that, at least on some RNAs or exon junctions, CASC3 assembly may occur soon after splicing within peripseckles as previously suggested (Daguenet et al., 2012).

CASC3 is a more prominent component of cytoplasmic EJCs within mRNPs that have not yet been translated or are undergoing their first round of translation (Figure 5). Previously described functions of CASC3 within translationally repressed neuronal transport granules (Macchi et al., 2003) and posterior-pole localized *oskar* mRNPs in *Drosophila* oocytes (van Eeden et al., 2001) further support CASC3 being a component of cytoplasmic pre-translation mRNPs. It remains to be tested if, like its active role in *oskar* mRNA localization and translation repression (van Eeden et al., 2001), CASC3 also plays a direct role in translational quiescence of RP mRNAs and/or their mobilization to rapidly meet increased demand for translation apparatus (Geyer et al., 1982; Patursky-Polischuk et al., 2009). Nonetheless, our results emphasize the idea of post-transcriptional regulons wherein mRNPs encoding functionally related activities are under coordinated translational control. Capture of such regulons via CASC3-EJC highlights a potential utility of EJC as a molecular marker to identify coordinately translated regulons, perhaps as a complementary approach to the recently described translation-dependent protein knock-off to monitor first round of translation of single mRNPs (Halstead et al., 2015).

**EJC composition and NMD**

Based on the binary EJC composition switch, EJC-dependent steps in the NMD pathway can be divided into at least two phases wherein the two alternate factors perform distinct functions. The susceptibility of all tested NMD targets to RNPS1 levels suggests that this protein, and perhaps other components of SR-rich EJCs, serve a critical function in an early phase in the pathway. Such a function could be to recruit and/or activate other EJC/UPF factors either in the nucleus or even during premature translation termination as part of the downstream EJC. Following the compositional switch, the EJC core maintains the ability to activate NMD as it can directly communicate with the NMD machinery via UPF3B (Buchwald et al., 2010; Chamieh et al., 2008). Still, following CASC3 incorporation into the EJC, its ability to stimulate NMD is likely reduced. Such a reduction may stem from the loss of RNPS1 or SR proteins, which are known to enhance NMD (Figure S7, (Aznarez et al., 2018; Sato et al., 2008; Viegas et al., 2007; Zhang and Krainer, 2004)). It is possible that CASC3 overexpression causes the compositional switch to occur at a faster rate or on a greater proportion of mRNAs, or both. Nevertheless, (over)expressed NMD reporters (e.g. β39 mRNA) and some endogenous mRNAs depend on both early and late EJC compositions for their NMD. Notably, the β-globin NMD reporter was previously shown to undergo biphasic decay with faster turnover around the nuclear periphery and slower decay in more distant cytoplasmic regions (Trcek et al., 2013). More recently, single-molecule imaging of reporter RNPs showed that a fraction of their population diffuses for several minutes and micrometers away from the nucleus before undergoing first round of translation (Halstead et al., 2015). It remains to be seen if mRNPs that are first translated in distant cytoplasmic locales, including those localized to specialized compartments such as neuronal dendrites and growth cones, may rely more on a CASC3-dependent slower phase of NMD. The EJC compositional switch may also underlie the distinct NMD branches identified earlier via tethering of RNPS1 and CASC3 to reporter mRNAs (Gehring et al., 2005). Consistent with these observations, we find that RNPS1 co-purifies more strongly with UPF2, while CASC3 appears to interact more with UPF3B (Figure 2 and data not shown). The nature of the relationship between EJC composition and UPF2 and UPF3B-independent NMD branches is an important avenue for the future work.

**ACKNOWLEDGEMENTS**

We would like to thank Pearlly Yan and the OSU Comprehensive Cancer Center genomics core for high-throughput sequencing, Philip Sharp and Paul Boutz for detained-intron datasets, Daniel Schoenberg for H1299 cell line Ribo-Seq and RNA-Seq datasets for translation efficiency estimation, Jens Lykke-Andersen for plasmids and antibodies, Akila Mayeda for anti-RNPS1 antibody, Can Cenik for advice, and Anita Hopper for comments on the manuscript. This work was supported in part by an allocation from the Ohio Supercomputer Center. Funding for this work was provided by the Ohio State University and National Institutes of Health (R01-GM120209) to G.S.

**AUTHOR CONTRIBUTIONS**

Conceptualization, G.S. and R.B.; Investigation, J.W.M., L.A.W., R.P., Z.Y., M.J., and G.S.; Writing – Original Draft, G.S., J.W.M. and L.A.W.; Writing – Review & Editing, G.S., J.W.M., L.A.W., R.P., Z.Y., M.J., V.W. and R.B.; Funding Acquisition, G.S.; Resources, G.S., V.W. and R.B.; Supervision, G.S., V.W. and R.B.

**DECLARATION OF INTERESTS**

The authors declare no competing interests.

**DATA AVAILABILITY**

All high-throughput DNA sequencing data are being submitted to GEO and will be available upon publication under Bioproject PRJNA471492.

# Figure 1

## A
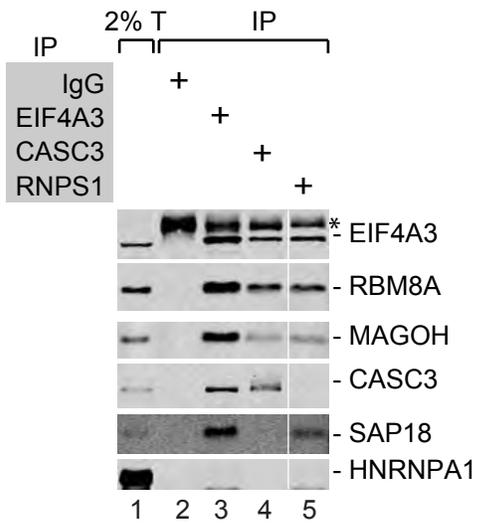

## B
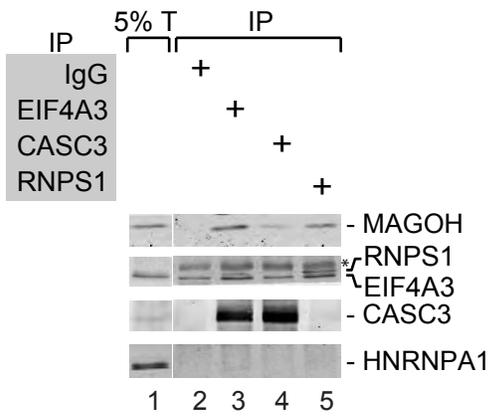

## C
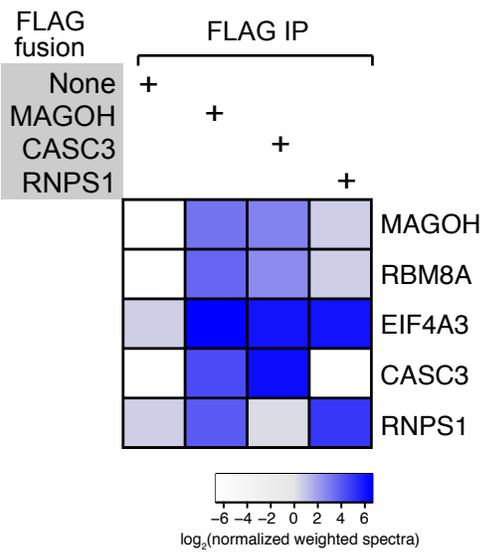

**Figure 1.** RNPS1 and CASC3 exists in mutually exclusive EJCs in mammalian cells.
- (A) Alternate EJCs in HEK293 cells. Western blots showing proteins on the right in total cell extract (T) (lane 1) or the immuno-precipitates (IP) (lanes 2-5) of the antibodies listed on the top. The asterisk (*) indicates signal from IgG heavy chain.
- (B) Alternate EJCs in mouse cortical total extracts and IPs as in (A).
- (C) Confirmation of mutually exclusive interactions via bottom-up proteomics. Heatmap showing signal for normalized-weighted spectra observed for the proteins on the right in the indicated FLAG IPs (top). The color scale for $\log_2$ transformed normalized-weighted spectral values is shown below.

# Figure 2

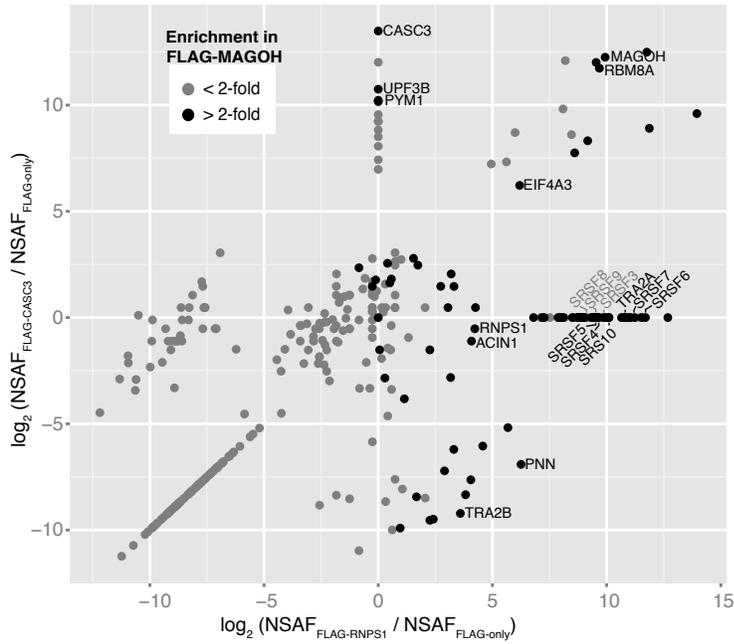
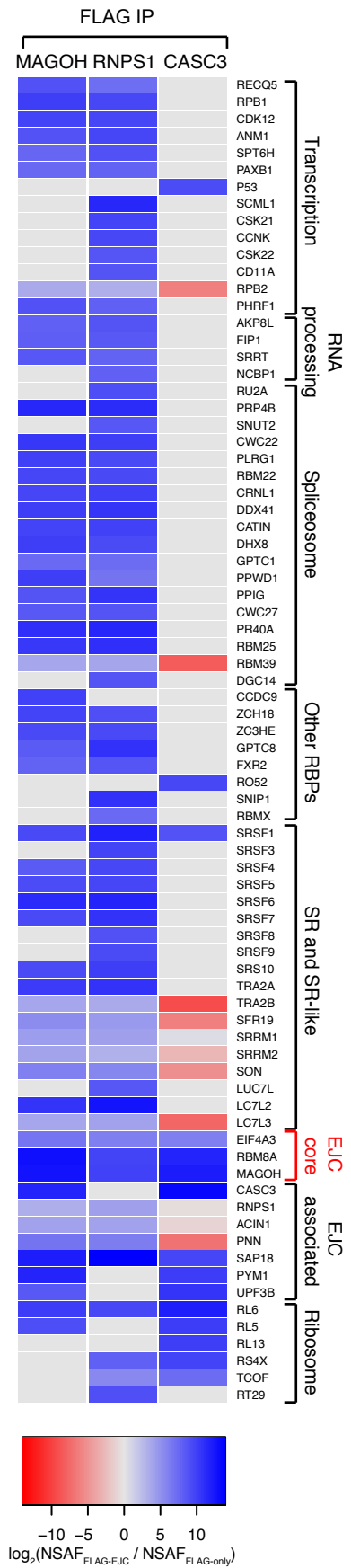
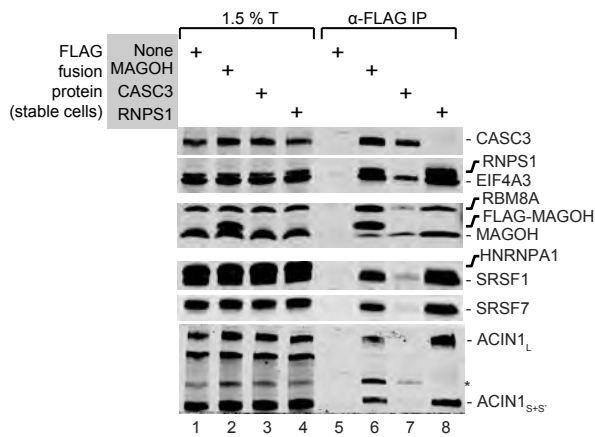
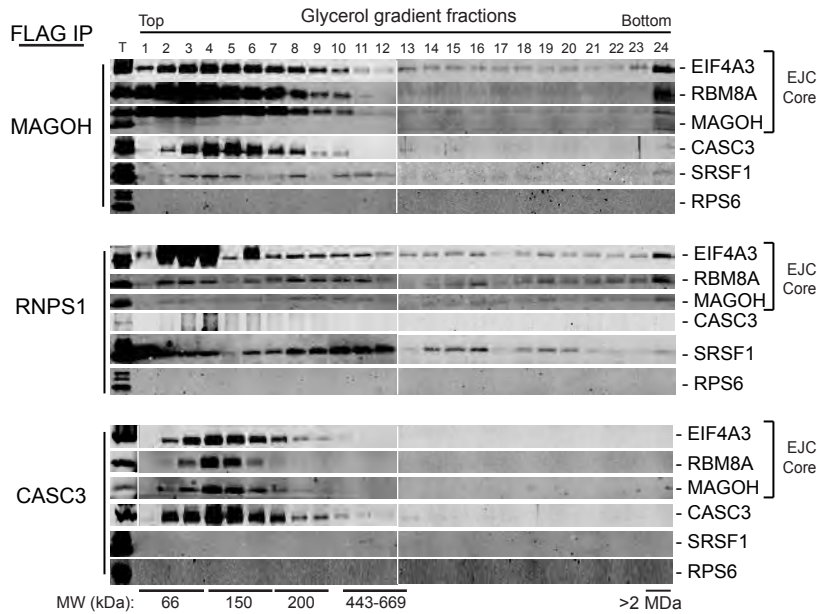

**Figure 2.** CASC3- and RNPS1-containing complexes have distinct protein composition and hydrodynamic size.

(A) Comparison of alternate EJC proteomes-I. A scatter plot where $\log_2$ transformed values are compared for fold-enrichment of proteins in FLAG-RNPS1 IP over FLAG-only control (x-axis) to fold-enrichment of the same proteins in FLAG-CASC3 IP over FLAG-only control (y-axis). Protein quantification was performed using Normalized Spectral Abundance Factor (NSAF). Each dot represents a protein that was identified in FLAG-EJC or FLAG-only control samples by Scaffold (see experimental procedures). Black dots: proteins enriched >2-fold over control in the FLAG-Magoh IP.

(B) Comparison of alternate EJC proteomes-II. A heatmap showing proteins that are >10-fold enriched in any of the FLAG-EJC IPs (indicated on the top) over FLAG-only control (proteins deemed to be contaminants were omitted). Proteins are grouped according to common functions (right). The color scale for the heatmap is at the bottom.

(C) Validation of mass spec. Western blots showing proteins (right) in total extract (T) or FLAG-IPs as on the top from HEK293 cells stably expressing FLAG-tagged proteins indicated on the top left. Note that RNPS1 migrates just above EIF4A3. The asterisk (*) indicates cross-reaction with CASC3 observed with anti-ACIN1 antibody.

(D) Hydrodynamic sizes of the alternate EJCs. Western blots showing proteins on the right in glycerol gradient fractions of FLAG-IPs from HEK293 cells stably expressing FLAG-MAGOH, FLAG-RNPS1, or FLAG-CASC3 (far left). The position of fractions from gradients (top) and molecular weight standards are indicated (bottom).

Figure 3

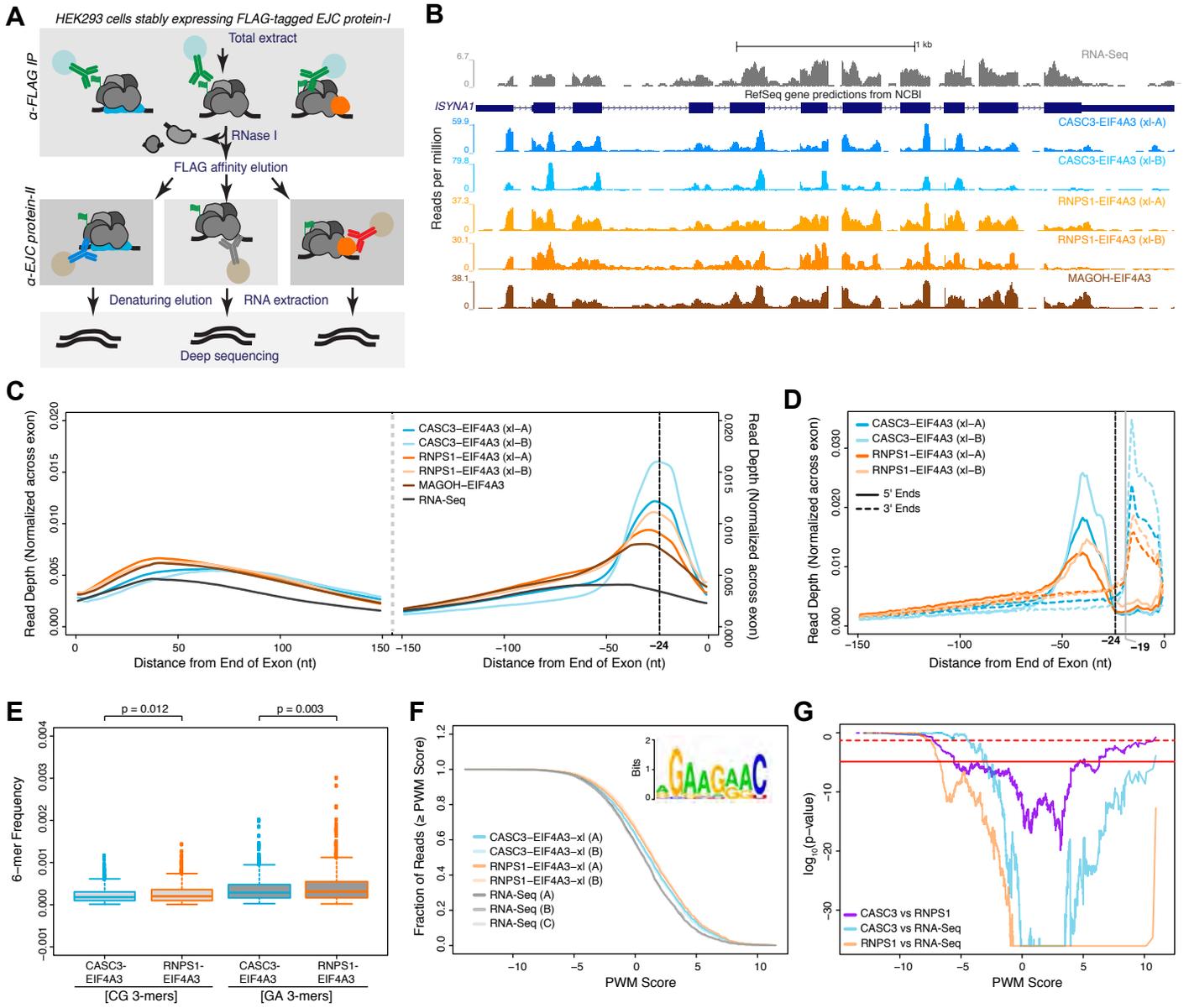

**Figure 3.** RNPS1 assembles with canonical and non-canonical EJCs but CASC3 is mainly part of canonical EJCs.

(A) A schematic illustrating main steps in RNA:protein Immunoprecipitation in Tandem (RIPiT)-Seq.

(B) Genome browser screen-shots comparing read coverage along the *ISYNA1* gene in RNA-Seq or RIPiT-Seq libraries (indicated on the right). The y-axis of each track in this window was normalized to millions of reads in each library. Blue rectangles: exons; thinner rectangles: untranslated regions; lines with arrows: introns.

(C) Meta-exon plots showing read depth in different RIPiT-Seq or RNA-Seq libraries (indicated in the middle) in the 150 nucleotides (nt) from the exon 5' end (left) or from the exon 3' end (right). The vertical black dotted line marks the -24 nt position previously determined to be the site of canonical EJCs (Singh et al., 2012).

(D) A composite plot of RNPS1 and CASC3 RIPiT-Seq footprint read 5' (solid lines) and 3' (dotted lines) ends. The canonical EJC site (-24 nt) is indicated by the vertical dotted line. Note there is some variability in the 5' boundary of EJCs whereas the 3' boundary is more strict (indicated by the vertical grey line at the -19 position).

(E) Box plots showing frequencies of 6-mers that contained CG 3-mers (CGG, GCG, CCG, CGC) or GA 3-mers (GGA, GAA, AGG, GAG) in CASC3 or RNPS1 RIPiT-Seq reads. Top: p-values based on Wilcoxon rank-sum test.

(F) Cumulative distribution function plots showing frequency of reads in the indicated samples with the highest score for match to SRSF1 motif (inset) position weight matrix (PWM). Sample identity is in the legend on the bottom left. The SRSF1 motif shown in the inset is from (Tacke and Manley, 1995).

(G) A negative binomial based assessment of significance of differences in SRSF1 motif PWM scores in (F) between the two alternate EJC footprint reads or between alternate EJC and RNA-Seq reads (legend on the bottom left). The horizontal dotted red line indicates p=0.05 whereas the horizontal solid red line indicates the Bonferroni adjusted p-value.

# Figure 4

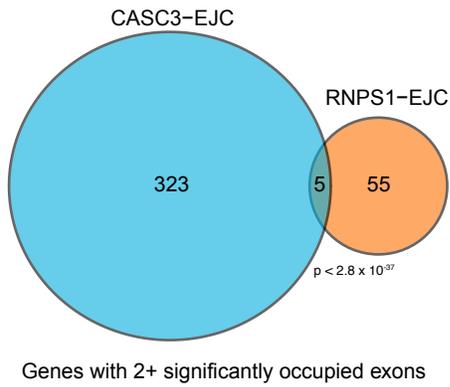
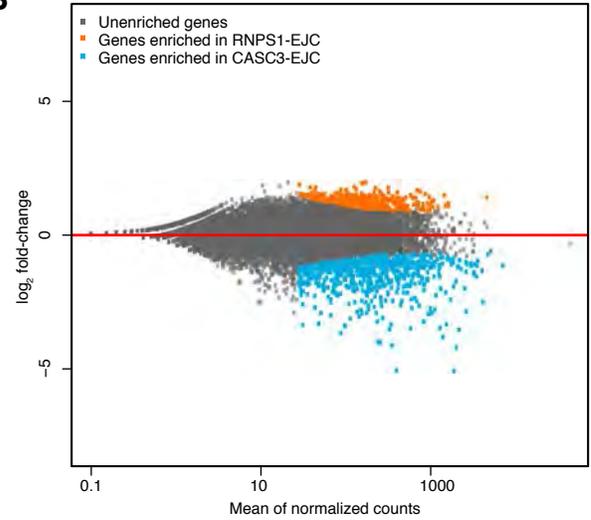
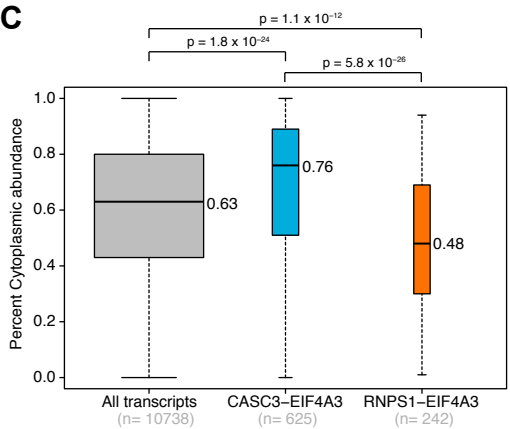
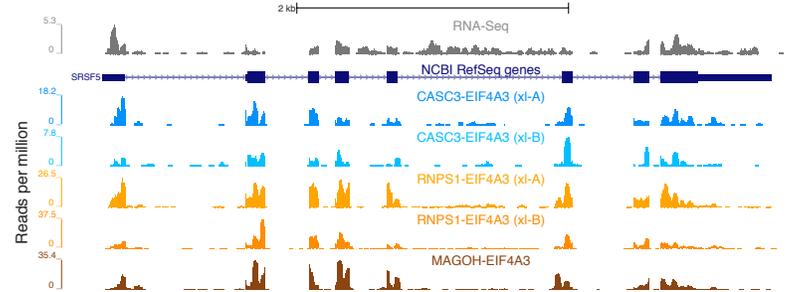
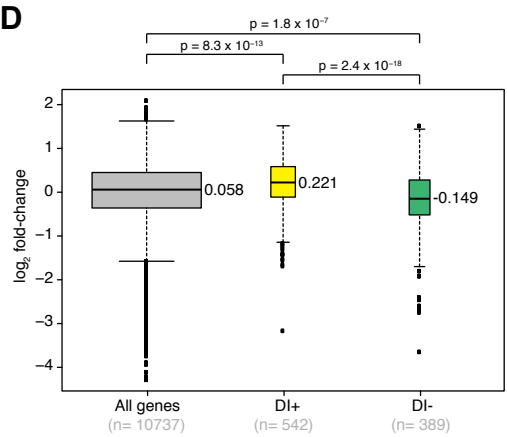
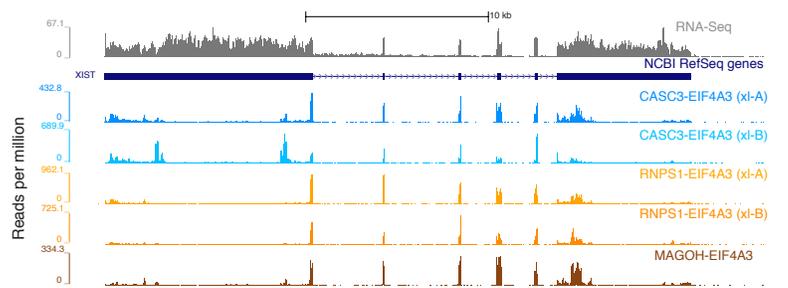

**Figure 4.** RNPS1 and CASC3 occupy same RNAs and sites but in different subcellular compartments.

(A) A venn-diagram showing genes with at least two differentially enriched exons, for which all differentially enriched exons are preferentially bound to CASC3-EJC (blue region), RNPS1-EJC (orange region), or genes with at least one exon is enriched in CASC3-EJC and one in RNPS1-EJC (overlap region). ($p < 2.8 \times 10^{-37}$). p-value shown was obtained as the probability of observing an overlap of 5 or fewer genes using a binomial distribution approximating each gene as containing two exons.

(B) An MA-plot showing fold-change in RNPS1-EJC versus CASC3-EJC footprint reads (y-axis) against expression levels (x-axis). Each dot represents a canonical transcript for each known gene in GRChg38 from UCSC "knownCanonical" splice variant table. Transcripts determined to be differentially enriched (p-adjusted<0.05) in RNPS1-EJC (orange) and CASC3-EJC (blue) are indicated.

(C) Box plots showing distribution of percent cytoplasmic levels (y-axis) for all (grey), CASC3-EJC enriched (blue), and RNPS1-EJC enriched (orange) transcripts. The median values are given to the right of each box plot. p-values on the top are based on Wilcoxon rank sum test. Bottom: The number of transcripts in each group.

(D) Box plots showing distribution of fold-change values (y-axis) for RNPS1-EJC versus CASC3-EJC footprint read counts for all (grey), detained intron-containing (DI+; yellow) or detained intron-lacking (DI-; green) transcripts. (+ve values: enriched in RNPS1-EJC; -ve values: enriched in CASC3-EJC). Top: p-values (Wilcoxon rank sum test). Bottom: the number of transcripts in each group.

(E) Genome browser screen-shots showing read coverage along the *SRSF5* gene in RNA-Seq (top) or RIPiT-Seq libraries (labeled on right). Blue rectangles: exons; thinner rectangles: untranslated regions; lines with arrows: introns. Note increased RNA-Seq reads in introns 4 and 5.

(F) Genome browser screen-shot as in (E) for the *XIST* locus.

Figure 5

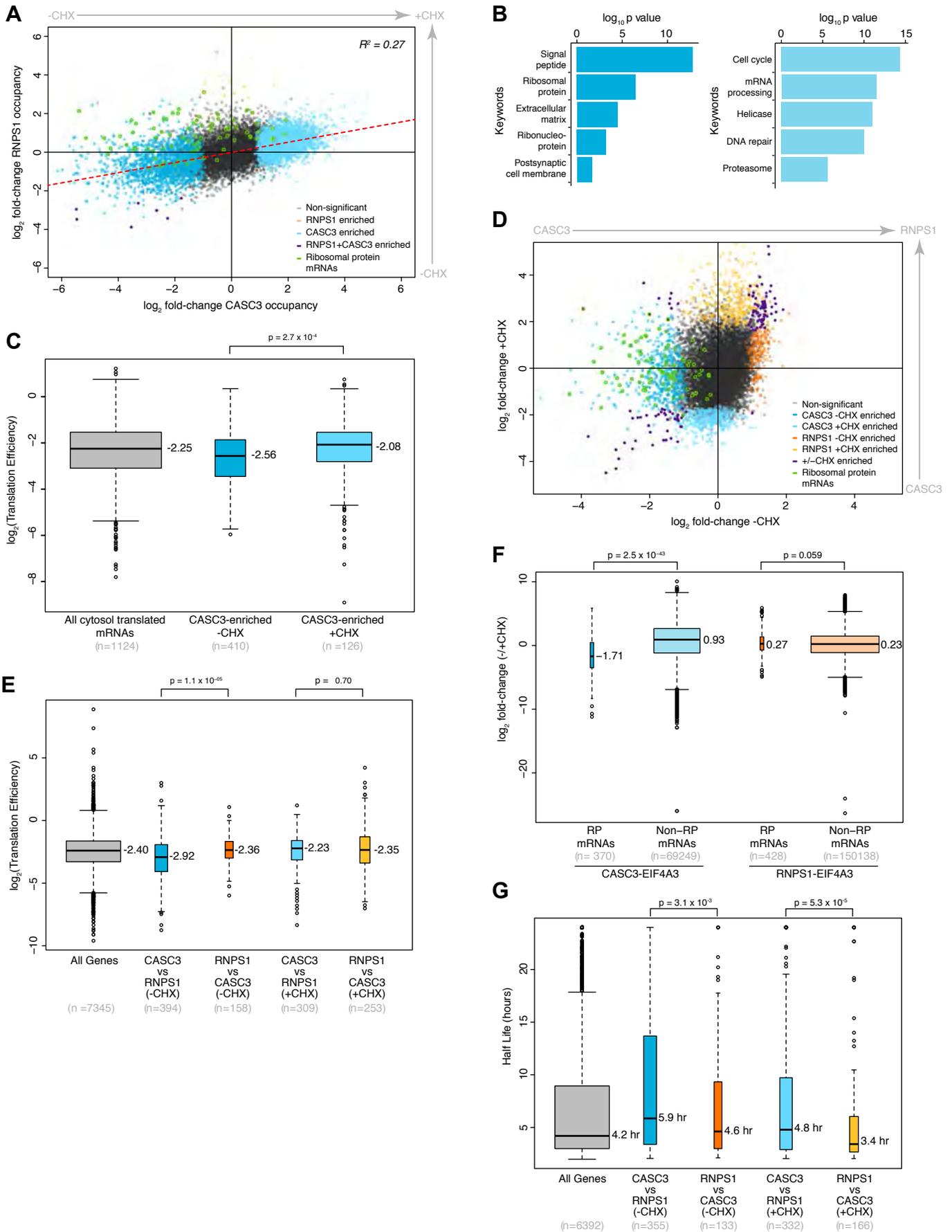

**Figure 5.** Translation and mRNA decay kinetics impacts alternate EJC occupancy.

(A) A scatter plot showing fold-change in CASC3-EJC occupancy in absence versus presence of cycloheximide (CHX; x-axis) and fold-change in RNPS1-EJC in absence versus presence of CHX (y-axis). Each dot represents a canonical transcript for each known gene in GRChg38, and is colored as indicated in the legend (bottom right). The dots representing ribosomal protein genes are highlighted by a green outline. The dotted red line shows a linear fit and the coefficient of determination ($R^2$) is on the top left corner.

(B) Top 5 GO term keywords and their $\log_{10}$ transformed p-values of enrichment in CASC3-EJC enriched transcripts (from A) in the absence (left) or presence of CHX (right).

(C) Box plots showing distribution of translation efficiency estimates (y-axis) of transcript groups on y-axis. The median values are given to the right of each box plot. Top: p-values (Wilcoxon rank sum test). Bottom: the number of transcripts in each group.

(D) Scatter plot as in (A) comparing fold-change in CASC3-EJC versus RNPS1-EJC footprint read counts in -CHX (x-axis) and +CHX (y-axis) conditions.

(E) Box plots as in (C) showing distribution of translation efficiency estimates of transcript groups from (D) as indicated on the bottom.

(F) Comparison of fold-change in CASC3-EJC or RNPS1-EJC footprint reads at canonical EJC sites (as defined in the methods section) from ribosomal protein (RP)-coding mRNAs or non-ribosomal (non-RP)-protein coding mRNAs. Top: p-values (Wilcoxon rank sum test).

(G) Box plots as in (C) and (E) above comparing mRNA half-life of transcript groups from (D) as indicated on the bottom.

# Figure 6

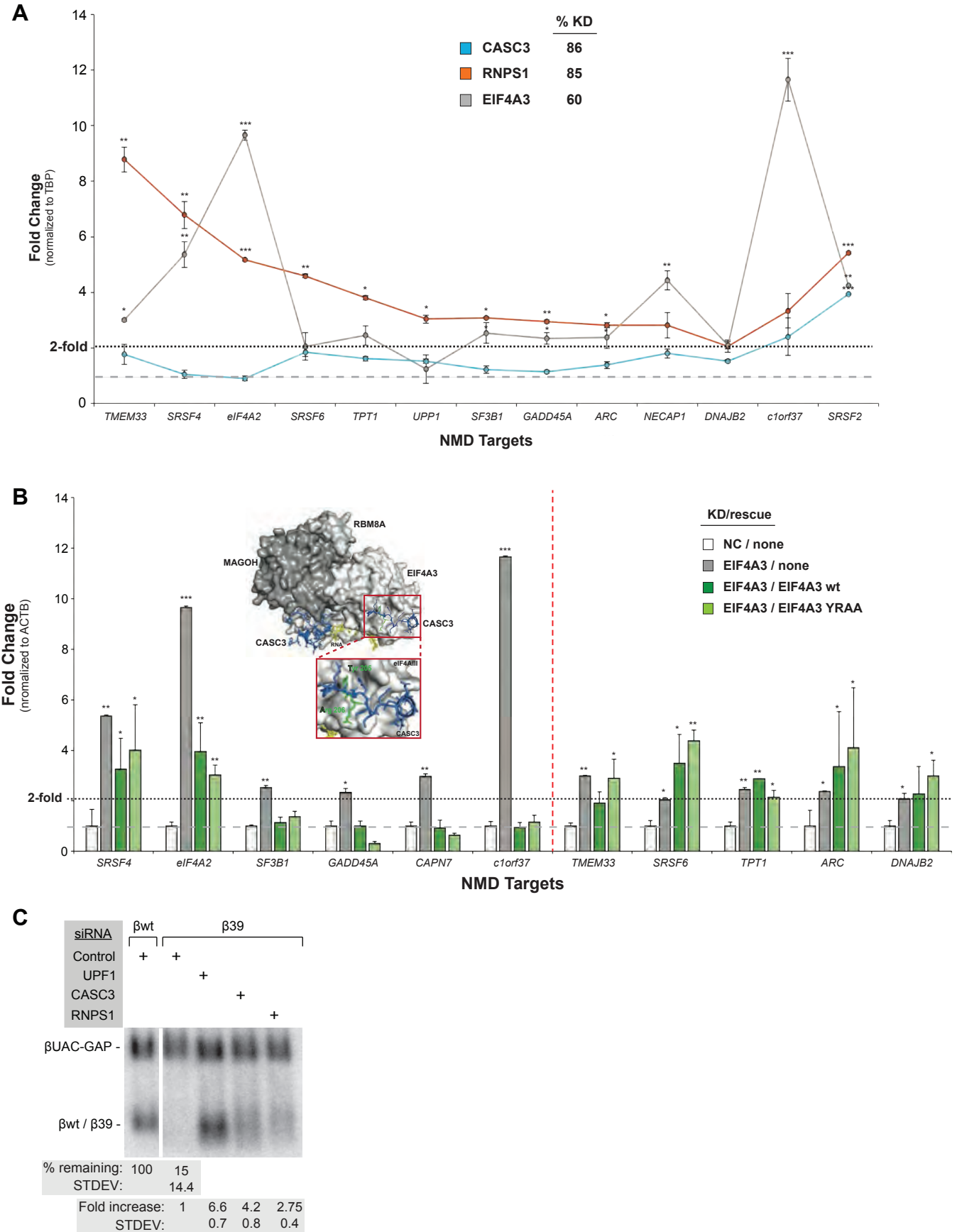

**Figure 6.** RNPS1 is required whereas CASC3 is largely dispensable for efficient NMD.

(A) Fold change as measured by real-time PCR in abundance of endogenous NMD-targeted transcripts (bottom) in HEK293 cells depleted of core or alternate EJC factors. (EIF4A3 knockdown: 48 hr, alternate EJC factor knockdown: 96 hr). Shown are the average values normalized to TBP levels from three biological replicates ± standard error of means (SEM). Asterisks denote statistically significant differences: * p<0.05, ** p<0.01, *** p<0.001 (Welch's t-test). Percent knockdown of each protein in a representative experiment is in the legend (see Figure S6).

(B) Real-time PCR analysis as in (A) from HEK293 cells depleted of EIF4A3. Either wild-type EIF4A3 or a mutant lacking CASC3 interaction were exogenously expressed in EIF4A3 knockdown cells as indicated in the legend on the top right. Inset: An EJC core crystal structure (PDB ID: 2J0S) showing the EIF4A3 residues that interact with CASC3 (enlarged below). The vertical red dotted line divides the NMD targets into two groups based on the degree of NMD restoration.

(C) Northern blots showing levels of wild-type (βwt, lane 1) or PTC-containing (codon 39; β39; lanes 2-5) β-globin mRNA and a longer internal control β-globin (βUAC-GAP) mRNA from HeLa Tet-off cells treated with siRNAs indicated on the top. Tables below indicate percentage of normalized β39 mRNA as compared to normalized βwt mRNA (top) or fold-increase in β39 mRNA upon knockdown as compared to control.

# Figure 7

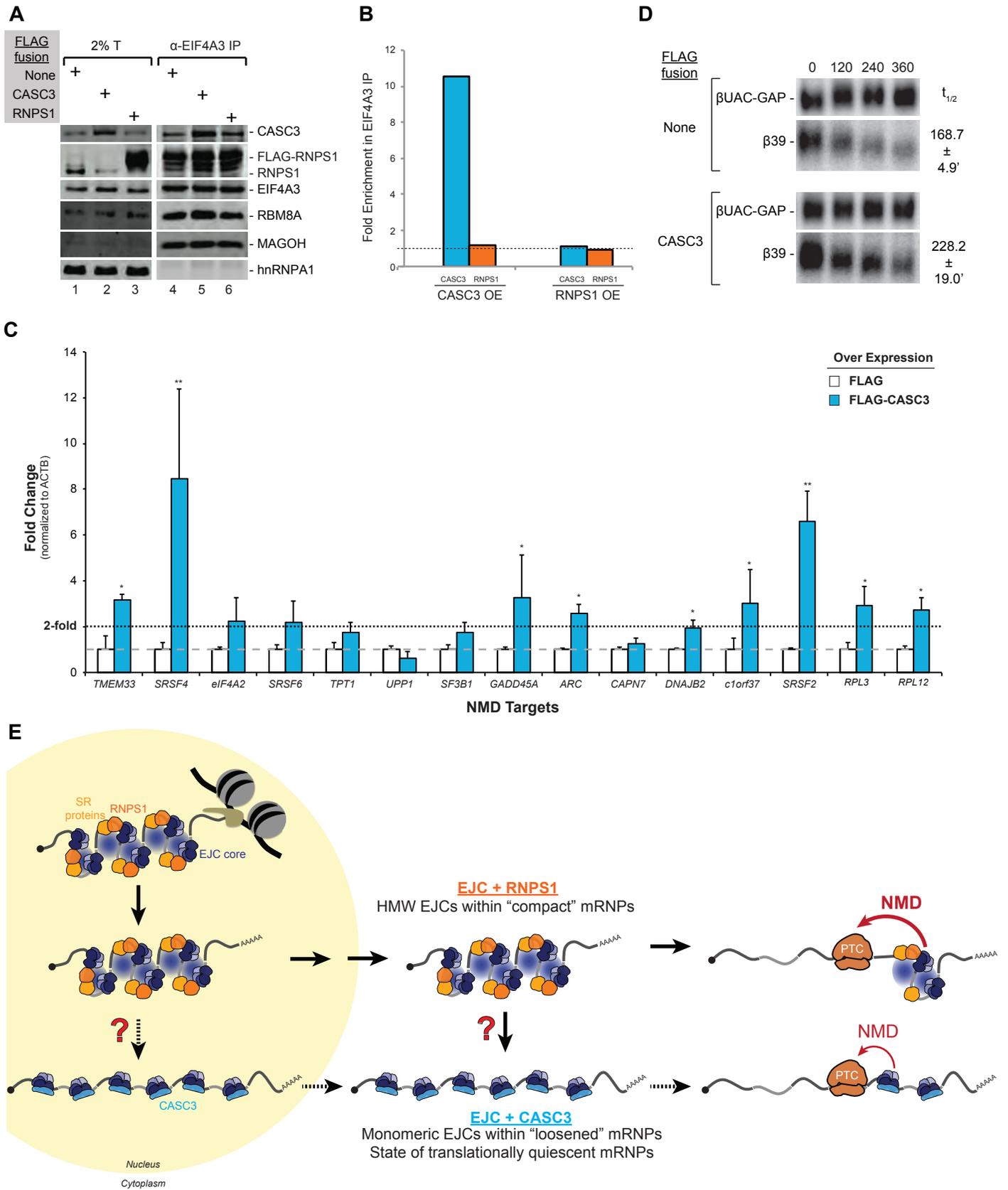

**Figure 7.** Promotion of switch to late-acting CASC3-EJC dampens NMD activity.

(A) Western blots showing proteins on the right in total extract (T, lanes 1-3) or EIF4AIII IPs (lanes 4-6) from HEK293 cells overexpressing FLAG-fusion proteins at top left.

(B) Histogram showing the fold enrichment of alternate EJC factors in EIF4A3 IPs from (A). Overexpressed (OE) alternate EJC factors are at the bottom and CASC3 (blue) and RNPS1 (orange) levels in each OE sample (lane 5 or 6) were compared to the control IP in lane 4 (grey, set to 1).

(C) Fold change in abundance of NMD-targeted endogenous transcripts (bottom) in HEK293 cells exogenously overexpressing FLAG or FLAG-CASC3. Shown are the average values normalized to ACTB levels from three biological replicates ± SEM. Asterisks denote statistically significant differences: * $p<0.05$, ** $p<0.01$, *** $p<0.001$ (Welch's t-test).

(D) Northern blots showing decay of Tetracycline-inducible β39 mRNA as compared to constitutively expressed βUAC-GAP mRNA in HeLa Tet-off cells overexpressing FLAG-tagged proteins indicated on the left. Time after Tet-mediated transcriptional shut-off of β39 mRNA is indicated on the top. On the right is β39 mRNA half-life ($t_{1/2}$) from each condition (average of three biological replicates ± standard deviation).

(E) A model depicting switch in EJC composition and its effect on mRNP structure and NMD activity. Key EJC proteins are indicated. Oval with radial blue fill in the high molecular weight (HMW) EJCs represent unknown interactions that mediate EJC multimerization. Grey and black lines: exons.

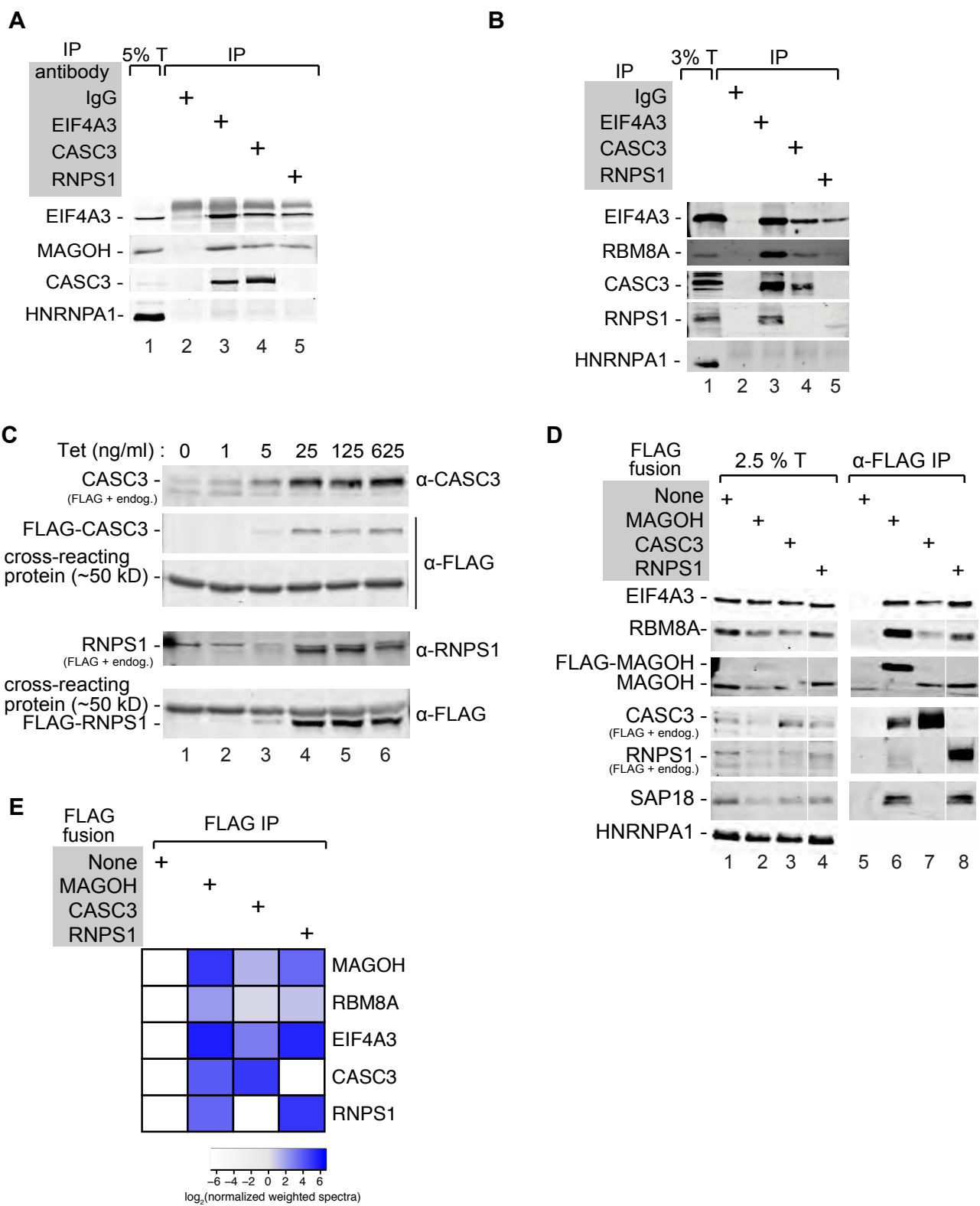

Figure S1

**Figure S1. (related to Figure 1)**

(A) Alternate EJCs in P19 cells. Western blots showing proteins on the left in P19 total cell extract (T, lane 1) or the immuno-precipitates (IP, lanes 2-5) using the antibodies listed on the top.

(B) Alternate EJCs in HeLa cells as in (A).

(C) Tetracycline (Tet)-induction of FLAG-tagged CASC3 and RNPS1. Western blots showing Tet-induced expression of FLAG-fusion proteins in FLP-In TREx Tet-on HEK293 cell lines. Tet concentrations are on the top, proteins detected are on the left and antibodies used for westerns are on the right.

(D) Tagged proteins assemble into alternate EJCs. Western blots showing proteins on the left in TE or FLAG IPs of FLAG-fusion proteins expressed in HEK293 cells at near-endogenous levels.

(E) Alternate EJC proteomic analysis (replicate 2). Heatmap showing enrichment of proteins on the right in FLAG-EJC protein IP proteomic analysis (indicated on the top).

# Figure S2

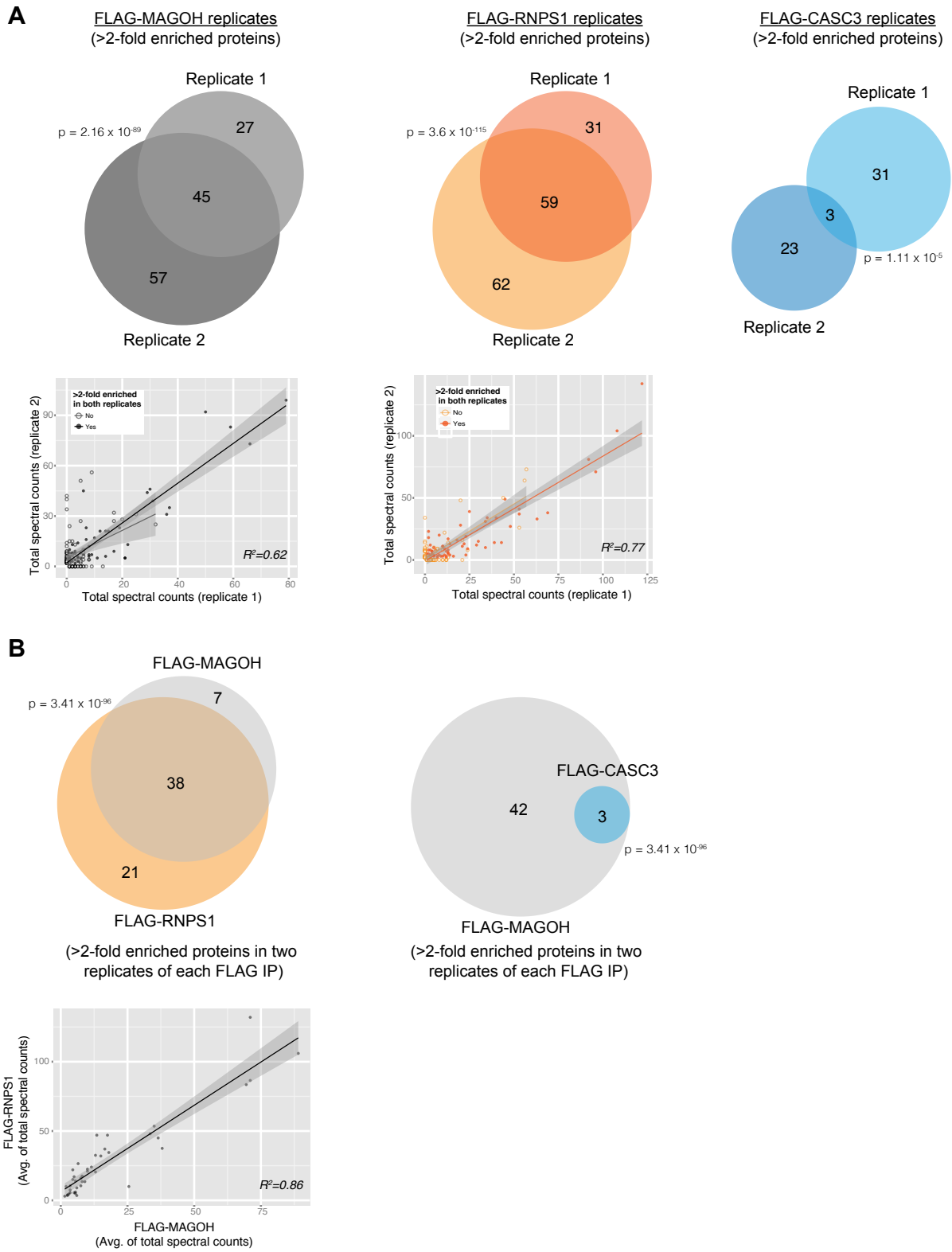

**Figure S2. (related to Figure 2)**

(A) Comparison between the two biological replicates of each FLAG-EJC mass spec samples. Overlap between >2-fold enriched proteins in the two biological replicates of proteomics samples of FLAG-EJC proteins indicated on the top. Hypergeometric test p-values are shown next to each venn diagram. For FLAG-MAGOH and FLAG-RNPS1 replicates, scatter plots compare total spectral counts for proteins detected in the two biological replicates. The linear fits for proteins enriched less or more than 2-fold are shown as indicated in the legend. Spearman correlation coefficient for >2-fold enriched proteins in each sample is shown.

(B) Comparisons as in (A) between the reproducibly enriched proteins in FLAG-MAGOH with those enriched in both replicates of FLAG-RNPS1 (left) and FLAG-CASC3 (right).

# Figure S3

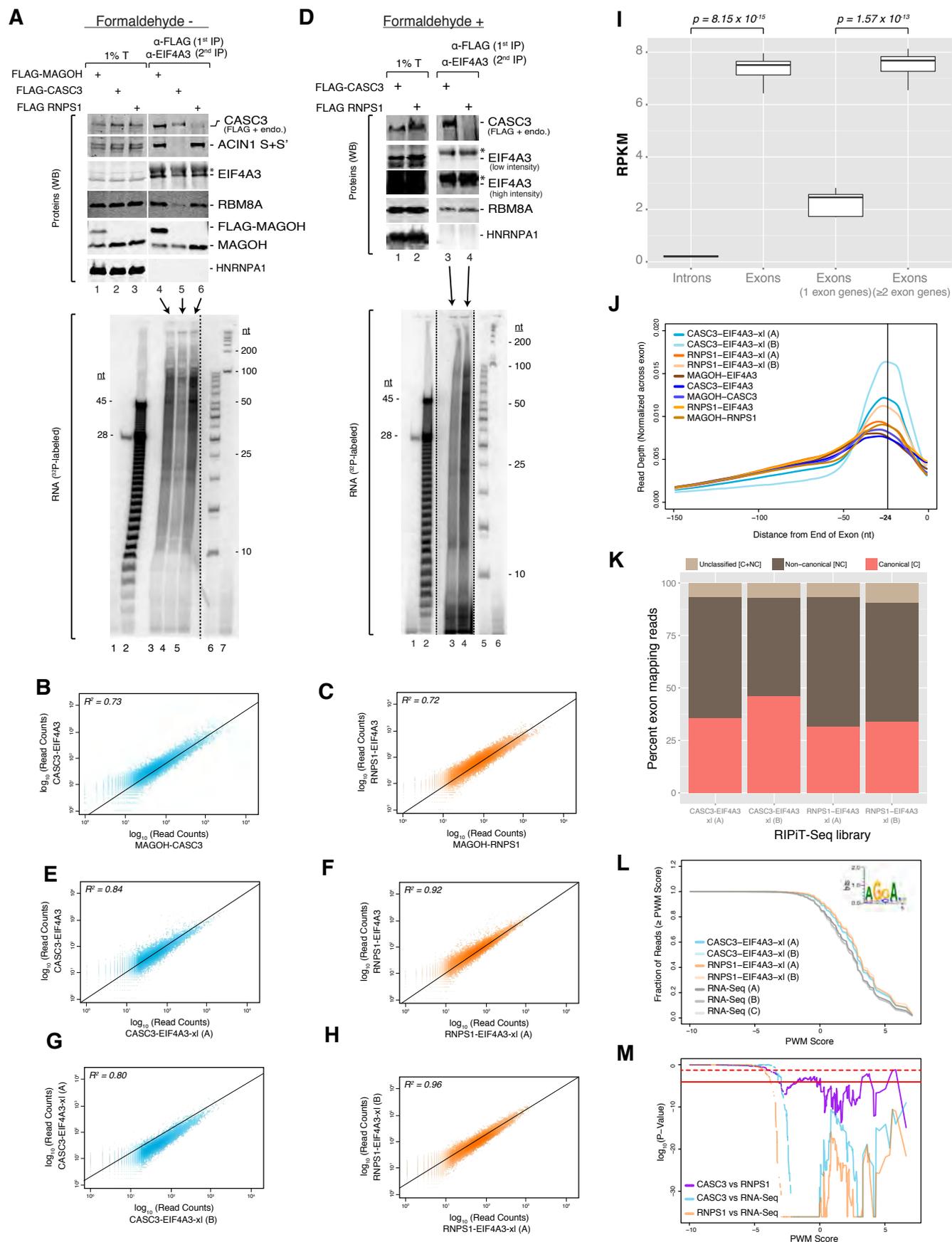

**Figure S3. (related to Figure 3)**

(A) Proteins and RNAs enriched in native alternate EJC RIPiTs. Top: Western blots with proteins indicated on the right in TE or RIPiT elutions. Bottom: Autoradiogram of end labeled RNAs from RIPiT elutions as well as size-markers (indicated on either side). The asterisk (*) indicates detection of IgG heavy chain.

(B) Comparison between native CASC3-EJC RIPiT-Seq samples [FLAG-MAGOH:CASC3 (x-axis) versus FLAG-CASC3:EIF4A3 (y-axis)]. Scatter plot comparing read counts for canonical transcripts for each known gene in GRChg38 (represented by each dot). A linear fit and Spearman correlation coefficient are also shown.

(C) Comparison between native RNPS1-EJC RIPiT-Seq samples [FLAG-MAGOH:RNPS1 (x-axis) versus FLAG-RNPS1:EIF4A3 (y-axis)] as in (C).

(D) Proteins and RNAs enriched in formaldehyde-crosslinked alternate EJC RIPiTs as in (A).

(E) Comparison as in (C) between native and formaldehyde-crosslinked CASC3-EJC RIPiT-Seq samples [FLAG-CASC3:EIF4A3, formaldehyde-crosslinked (xl), replicate A (x-axis) versus native FLAG-CASC3:EIF4A3 (y-axis)].

(F) Comparison as in (C) between native and formaldehyde-crosslinked RNPS1-EJC RIPiT-Seq samples [FLAG-RNPS1:EIF4A3, formaldehyde-crosslinked (xl), replicate A (x-axis) versus native FLAG-RNPS1:EIF4A3 (y-axis)].

(G) Comparison as in (C) between two replicates of the formaldehyde-crosslinked CASC3-EJC RIPiT-Seq samples [FLAG-CASC3:EIF4A3-xl, replicate B (x-axis) versus FLAG-CASC3:EIF4A3-xl, replicate A (y-axis)].

(H) Comparison as in (C) between two replicates of the formaldehyde-crosslinked RNPS1-EJC RIPiT-Seq samples [FLAG-RNPS1:EIF4A3-xl, replicate B (x-axis) versus FLAG-RNPS1:EIF4A3-xl, replicate A (y-axis)].

(I) Normalized read densities in alternate EJC footprint libraries in the genomic region indicated at the bottom. Each box plot comprises values from all eight RIPiT-Seq datasets listed in Table S2. Top: p-values (Wilcoxon rank sum test).

(J) Meta-exon plots showing read depth in different RIPiT-Seq or RNA-Seq libraries (top left corner) in the 150 nucleotides (nt) from the exon 3' end.

(K) Distribution of exon mapping reads among the exonic regions indicated on top in the RIPiT-Seq libraries (bottom).

(L) Cumulative distribution function plots showing frequency of reads in the indicated samples with the highest score for match to PWM for SRSF9 motif (inset, top right). Bottom left: sample identity. The SRSF9 motif shown in the inset is from (Paradis et al., 2007).

(M) A negative binomial based assessment of significance of differences in SRSF9 motif PWM scores in (L) between reads from samples in the legend on the bottom left. Horizontal dotted red line: p=0.05; Horizontal solid red line: Bonferroni adjusted p-value.

# Figure S4

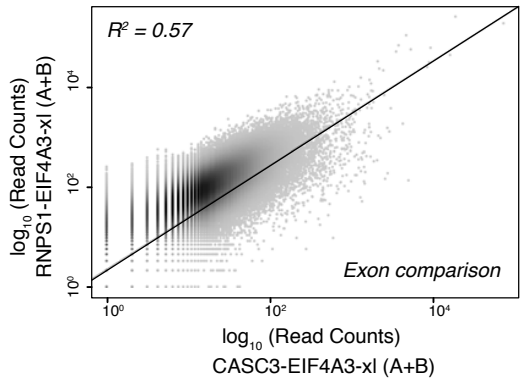

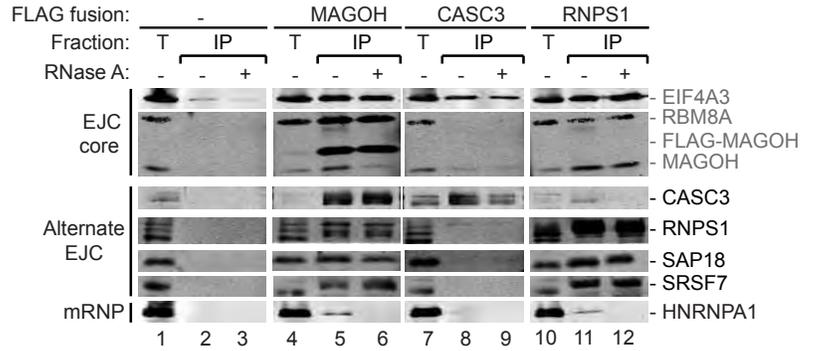

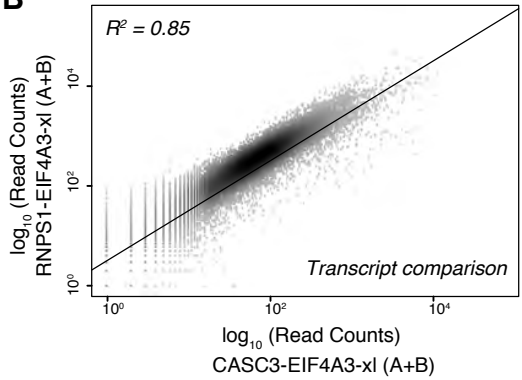

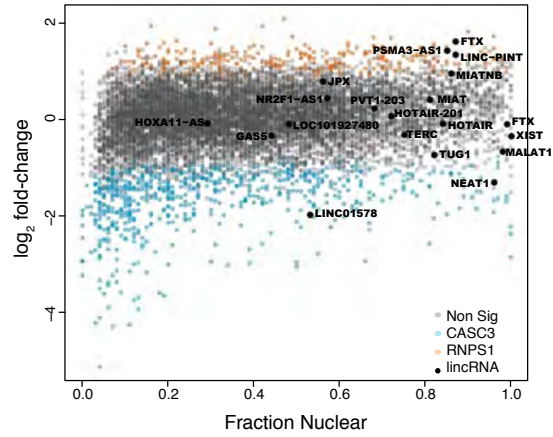

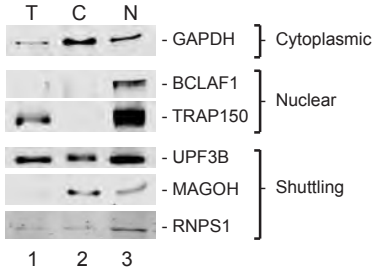

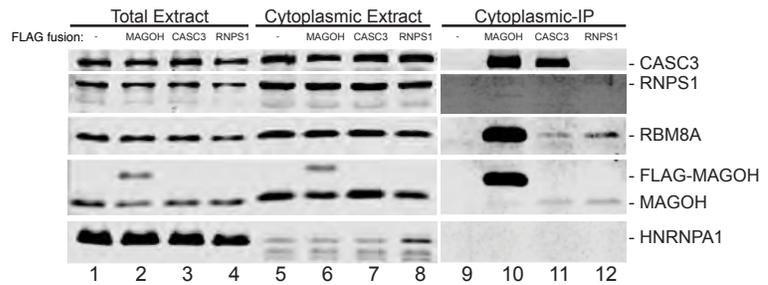

**Figure S4. (related to Figure 4)**

(A) Comparison of read counts from both cross-linked replicates of CASC3-EJC (x-axis) versus RNPS1-EJC (y-axis). Each dot on the scatter plot represents each exon in the knownCanonical transcript annotation for UCSC known genes. A linear fit and coefficient of determination ($R^2$) are shown.

(B) Comparison of read counts from both cross-linked replicates of CASC3-EJC (x-axis) versus RNPS1-EJC (y-axis). Each dot on the scatter plot represents knownCanonical transcript for UCSC known genes. A linear fit and coefficient of determination ($R^2$) are shown.

(C) RNA dependent and independent interactions between EJC proteins. Western blots of proteins on the right in total extracts (T) or FLAG immunoprecipitates (IP) from HEK293 cells expressing FLAG-tagged proteins on the top. Inclusion of RNase A in extracts during IP is indicated on top of each lane.

(D) Relative alternate EJC levels versus nuclear abundance of long non-coding RNAs (lncRNAs). Each dot on the scatter plot represents knownCanonical transcript with spliced lncRNAs labeled in black. CASC3 or RNPS1-enriched transcripts are colored as in the legend (bottom right). x-axis: -ve fold-change= CASC3-enrichment, +ve fold-change= RNPS1-enrichment.

(E) Subcellular fractionation. Western blots for proteins on the right in total extracts (T), cytoplasmic (C), or nuclear (N) fractions of HEK293 cells.

(F) Presence of alternate EJCs in cytoplasmic extracts. Western blots for proteins on the right in the fractions of HEK293 cells (top). FLAG-tagged proteins expressed in cells are indicated above each lane.

# Figure S5

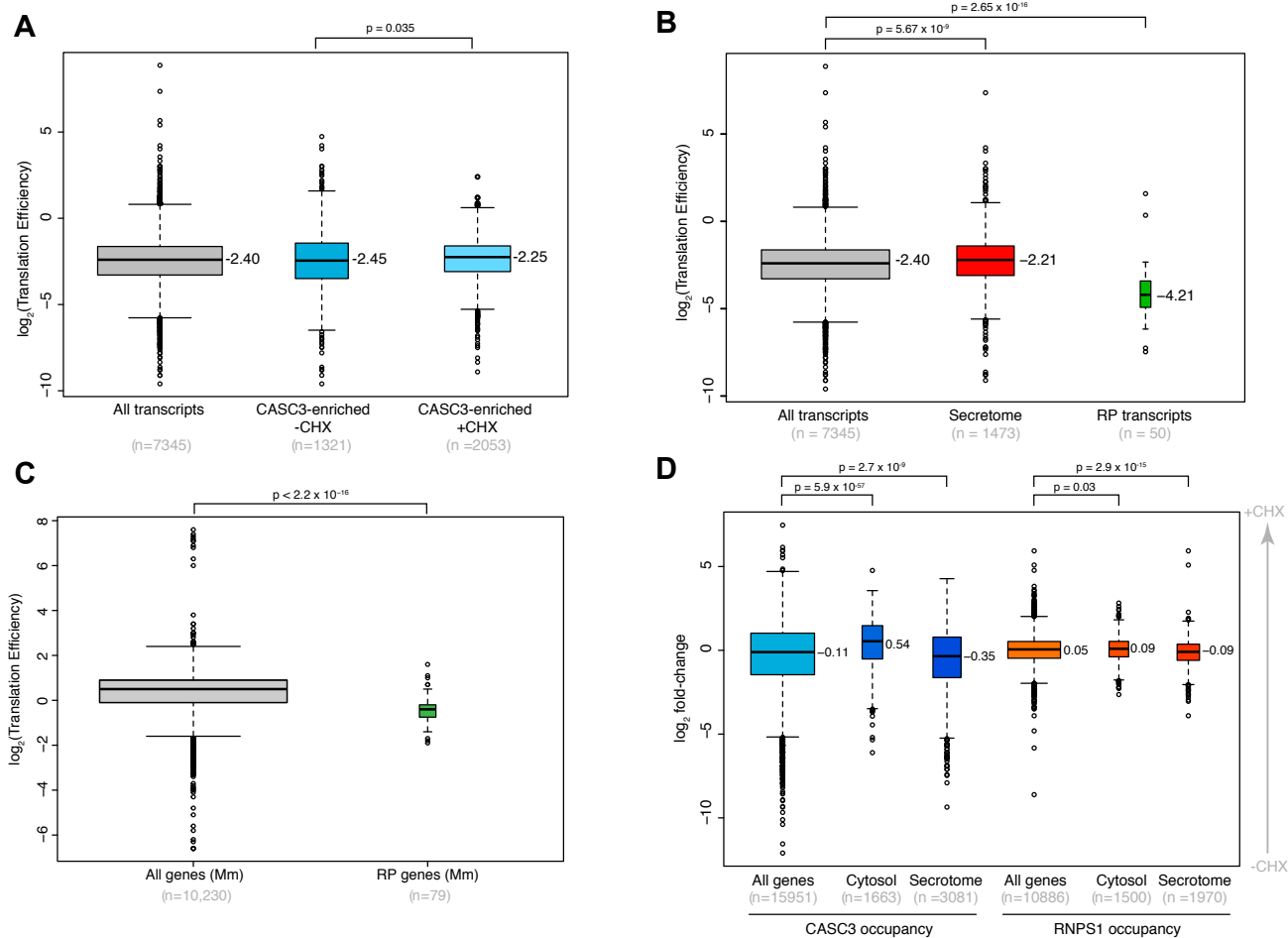

**Figure S5. (related to Figure 5)**

(A) Box plots showing distribution of translation efficiency estimates (y-axis) of transcript groups on y-axis. The median values are given to the right of each box plot. Top: p-values (Wilcoxon rank sum test). Bottom: the number of transcripts in each group.

(B) Comparison of translation efficiency (TE, y-axis) of secrotome (red) and ribosomal protein (RP) transcripts (green) to all transcripts (grey). Number of genes in group are at the bottom. TE data from (Kiss et al., 2017).

(C) Translation efficiency of RP genes in mouse embryonic stem cells. TE data from (Ingolia et al., 2011).

(D) Box plots showing fold-change (y-axis) in alternate EJC occupancy (x-axis, at the bottom) for transcript groups shown at the bottom. x-axis: -ve fold-change= enrichment in -CHX condition, +ve fold-change= enrichment in +CHX condition. The median fold-change values are given to the right of each box plot. Top: p-values (Wilcoxon rank sum test). Bottom: the number of transcripts in each group.

# Figure S6

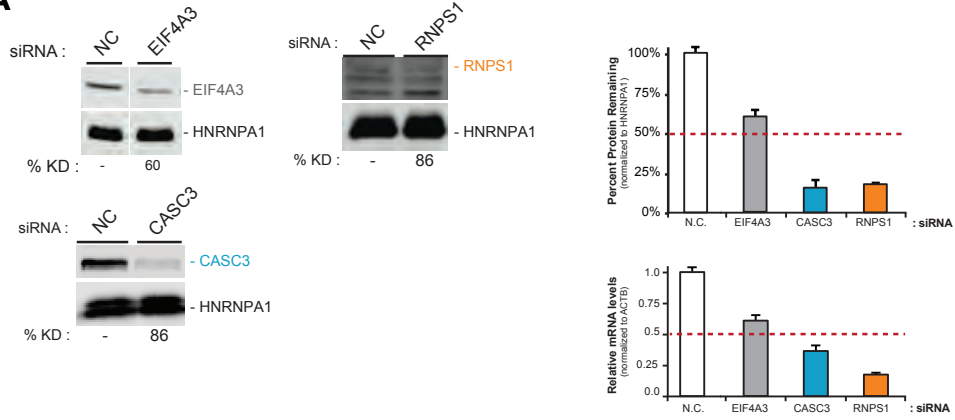
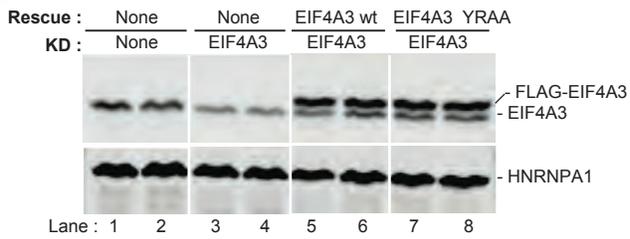
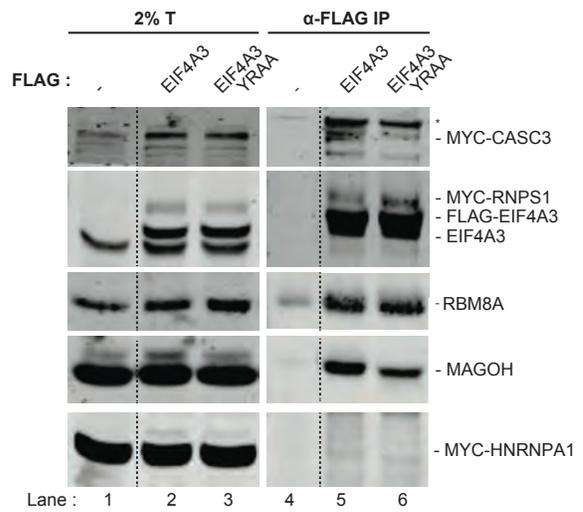
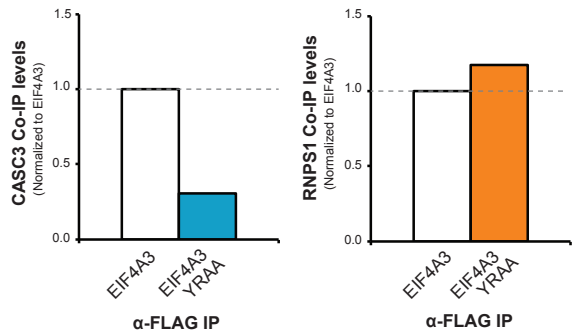
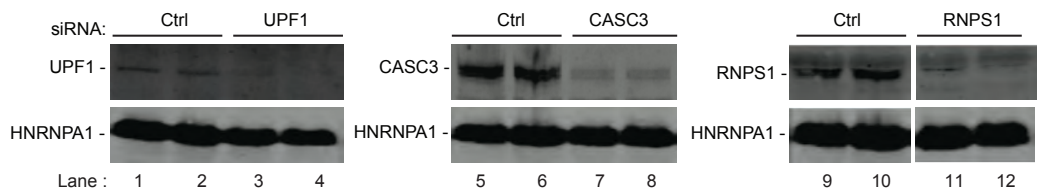

**Figure S6 (related to Figure 6)**

(A) Knockdowns of EJC proteins (related to Figure 6A). Western blots of EJC core and alternate factor siRNA-mediated knockdowns (left). Right, top: Quantification of knockdown efficiency of EJC protein remaining relative to HNRNPA1 protein levels. Right, bottom: Knockdown efficiency as determined by qPCR of EJC factor mRNA levels relative to ACTB mRNA levels .

(B) Rescue of EIF4A3-depleted cells. Western blots for proteins on the right from HEK293 cells depleted of EIF4A3 and expressing either wild-type EIF4A3 or a mutant EIF4A3 lacking CASC3 interaction as indicated on the top.

(C) Interaction of EIF4A3-YRAA with CASC3. Western blots showing proteins on the right in total extract or FLAG IPs from HEK293 cells expressing FLAG-tagged proteins on the top along with MYC-CASC3, -RNPS1, and -HNRNPA1. The asterisk indicates cross-reacting band likely to be unreduced FLAG-IgG heavy chain.

(D) Quantification of western blots in (C) showing relative MYC-CASC3 (left, blue) and MYC-RNPS1 (right, orange) co-IP with FLAG-EIF4A3 proteins on the bottom.

(E) Knockdown of EJC/UPF1 proteins (Figure 6C). Western blots showing proteins indicated on the left from HeLa Tet-off cells treated with siRNAs indicated on the top. Two technical replicates of each knockdown are shown. HNRNPA1 panel shown in lanes 5-6 is shown again in lanes 9-10.

# Figure S7

## A

|  | 1.5 % T | | | IP | | | |
|---|---|---|---|---|---|---|---|
|  |  |  |  | IgG | α-RBM8A | | |
| OE : | FLAG | CASC3 | RNPS1 | FLAG | FLAG | CASC3 | RNPS1 |

- CASC3
- RNPS1
- RBM8A
- MAGOH
- HNRNPA1

lanes 1 2 3 4 5 6 7

FOE :  -   5.0   15.2

## B

Fold Enrichment of CASC3 in α-RBM8A IP — FLAG, CASC3, RNPS1

Fold Enrichment of RNPS1 in α-RBM8A IP — FLAG, CASC3, RNPS1

Over Expression

## C

| FLAG fusion | βwt | | | β39 | | |
|---|---|---|---|---|---|---|
| None | + |  |  | + |  |  |
| RNPS1 |  | + |  |  | + |  |
| CASC3 |  |  | + |  |  | + |

βUAC-GAP  
βwt / β39

% Remaining :  100  100  100  12.3  7.5  15.3

| FLAG fusion | βwt | | | β39 | | |
|---|---|---|---|---|---|---|
| None | + |  |  | + |  |  |
| RNPS1 |  | + |  |  | + |  |
| CASC3 |  |  | + |  |  | + |

βUAC-GAP  
βwt / β39

% Remaining :  100  100  100  8.0  5.3  16.9

**Figure S7 (related to Figure 7)**

(A) Effect of alternate EJC factor overexpression on EJC composition. Western blots for proteins on the right in total extract (lanes 1-3), control IgG IP (lane 4) or RBM8A IPs (lanes 5-7) from HEK293 cells overexpressing (OE) FLAG-fusion proteins at top. Fold overexpression (FOE) on bottom is relative to protein levels in lane 1 after normalizing to HNRNPA1 protein levels in particular lane.

(B) Quantification of (A) for co-IP of CASC3 (left) or RNPS1 (right) upon RBM8A IP from cell lines overexpressing the proteins on the bottom. Fold enrichment was normalized by co-purifying MAGOH levels.

(C) Effect of alternate EJC overexpression on β-globin NMD. Northern blots showing levels of wild-type β-globin RNA (βwt) and an NMD-targeted β-globin RNA (β39) (as indicated on the top) along with co-transfected βUAC-GAP RNA from HeLa cells expressing alternate EJC factors (top left). % Remaining: βwt or β39 levels in each sample were normalized to βUAC-GAP RNA. Numbers below each lane show normalized β39 RNA remaining as % of normalized βwt RNA for the corresponding treatment (set to 100). Two independent replicates are shown.

**Table S1.** Summary of proteins identified by LC-MS/MS in the two replicates of FLAG-EJC and FLAG-only IPs.

| #Alternate EJC_replicate1 | | | |
|---|---|---|---|
| 341 proteins at 99% minimum, 2 minimum peptides, 0.0% decoy FDR;; 9540 spectra at 95% minimum, 0.00% decoy FDR | | | |
| From proteins clusters, only member proteins with a non-zero value in at least one sample were retained | | | |
| Normalized spectral abundance factor (NSAF) values from Scaffold were used for quantitation; a psuedo count of 0.000001 was added to each NSAF value before determining enrichment in EJC IPs as compared to FLAG-only | | | |
| Proteins enriched >2-fold in any of the three EJC IPs over FLAG-only IP are shown | | | |
| | | | |
| | | log2(NSAF) | |
| Accession Number | MAGOH | CASC3 | RNPS1 |
| SAP18_HUMAN | 12.12344246 | 9.600730633 | 13.95255926 |
| LC7L2_HUMAN | 10.76661172 | 0 | 12.67785175 |
| SRSF1_HUMAN | 9.42840205 | 8.905056005 | 11.86905398 |
| H4_HUMAN | 12.27926234 | 12.49192823 | 11.75826486 |
| SRSF6_HUMAN | 11.27676417 | 0 | 11.69657638 |
| PR40A_HUMAN | 11.06379952 | 0 | 11.54245153 |
| SCML1_HUMAN | 0 | 0 | 11.46142829 |
| PRP4B_HUMAN | 11.4926043 | 0 | 11.23463783 |
| RBM25_HUMAN | 10.46954007 | 0 | 11.04753294 |
| GPTC8_HUMAN | 8.417008587 | 0 | 10.96275144 |
| SNIP1_HUMAN | 0 | 0 | 10.95325069 |
| SRSF7_HUMAN | 9.487699374 | 0 | 10.81338039 |
| TRA2A_HUMAN | 10.24210228 | 0 | 10.79100005 |
| PPIG_HUMAN | 8.82527683 | 0 | 10.73479428 |
| DDX41_HUMAN | 10.10105646 | 0 | 10.64988486 |
| CWC22_HUMAN | 10.55487659 | 0 | 10.10446771 |
| SRS10_HUMAN | 9.349303347 | 0 | 10.09037711 |
| CRNL1_HUMAN | 9.654385415 | 0 | 9.980853606 |
| MGN_HUMAN | 12.77585716 | 12.25168951 | 9.934059395 |
| CATIN_HUMAN | 9.81605565 | 0 | 9.879828273 |
| CSK21_HUMAN | 0 | 0 | 9.834897633 |
| SRSF3_HUMAN | 0 | 0 | 9.766512345 |
| RBM8A_HUMAN | 12.93774162 | 11.73564064 | 9.681238412 |
| ANM1_HUMAN | 8.887708033 | 0 | 9.628390857 |
| SRSF5_HUMAN | 9.295332732 | 0 | 9.621740545 |
| CDK12_HUMAN | 9.840982435 | 0 | 9.583195621 |
| RPB1_HUMAN | 10.1382718 | 0 | 9.572643389 |
| RL6_HUMAN | 10.21176665 | 12.0086037 | 9.539391485 |
| PGAM5_HUMAN | 9.208014828 | 0 | 9.534400161 |
| CCNK_HUMAN | 0 | 0 | 9.529430554 |
| SRSF4_HUMAN | 8.43629512 | 0 | 9.497971558 |
| PLRG1_HUMAN | 9.961290512 | 0 | 9.440786132 |
| RBM22_HUMAN | 9.668034004 | 0 | 9.410281737 |
| ZC3HE_HUMAN | 9.443979543 | 0 | 9.408329741 |
| SRSF9_HUMAN | 0 | 0 | 9.336752009 |
| DHX8_HUMAN | 10.1291542 | 0 | 9.331432034 |
| H2AY_HUMAN | 9.842915995 | 8.321612471 | 9.170726276 |
| RU2A_HUMAN | 0 | 0 | 9.130647779 |
| H2A1A_HUMAN | 0 | 0 | 9.091646948 |
| RT29_HUMAN | 0 | 0 | 9.073445375 |
| ZCH18_HUMAN | 9.807709112 | 0 | 9.036201092 |
| SRSF8_HUMAN | 0 | 0 | 8.985699652 |
| SPT6H_HUMAN | 7.634593268 | 0 | 8.957073013 |
| NKTR_HUMAN | 9.675904121 | 0 | 8.933572638 |
| CD11A_HUMAN | 0 | 0 | 8.834629058 |
| CWC27_HUMAN | 8.501837185 | 0 | 8.82791427 |
| DGC14_HUMAN | 0 | 0 | 8.815767536 |
| AKP8L_HUMAN | 8.050556113 | 0 | 8.790283326 |
| RBBP6_HUMAN | 6.588219912 | 0 | 8.77778133 |
| FXR2_HUMAN | 7.991748578 | 0 | 8.731352976 |
| CSK22_HUMAN | 0 | 0 | 8.674721924 |
| HP1B3_HUMAN | 9.271719795 | 7.75181182 | 8.599950025 |
| LUC7L_HUMAN | 0 | 0 | 8.590886411 |
| SNUT2_HUMAN | 0 | 0 | 8.569058109 |
| FIP1_HUMAN | 8.171226847 | 0 | 8.497053445 |
| CMBL_HUMAN | 0 | 12.08988078 | 8.190713617 |
| SMU1_HUMAN | 9.964095343 | 0 | 8.124793379 |
| RS4X_HUMAN | 0 | 9.818933622 | 8.088788239 |
| NCBP1_HUMAN | 0 | 0 | 8.0869854 |
| PHRF1_HUMAN | 9.017838416 | 0 | 8.025527183 |
| SRRT_HUMAN | 8.609437329 | 0 | 7.938462236 |
| PAXB1_HUMAN | 7.547434405 | 0 | 7.87276721 |
| RBMX_HUMAN | 0 | 0 | 7.519243094 |
| GPTC1_HUMAN | 7.525755693 | 0 | 7.269126679 |
| RECQ5_HUMAN | 9.015526561 | 0 | 7.179610571 |
| PPWD1_HUMAN | 10.04644195 | 0 | 6.799993888 |
| PININ_HUMAN | 6.8673278 | -6.910133013 | 6.25017964 |
| IF4A3_HUMAN | 6.771047915 | 6.220329511 | 6.19106795 |
| DDB1_HUMAN | 0 | 8.704906538 | 5.99047842 |
| SON_HUMAN | 6.126682474 | -5.179948888 | 5.676165346 |
| TCOF_HUMAN | 0 | 7.326069892 | 5.61303063 |
| SFR19_HUMAN | 5.367405791 | -6.048388784 | 4.575401278 |
| SRRM1_HUMAN | 4.360907368 | 0.470829819 | 4.257178256 |
| RNPS1_HUMAN | 3.31613842 | -0.526593813 | 4.227919599 |
| ACINU_HUMAN | 4.051882067 | -1.108281217 | 4.080682378 |
| LC7L3_HUMAN | 3.797146519 | -7.636334602 | 4.053576036 |
| RBM39_HUMAN | 3.799733114 | -8.33860269 | 3.82174148 |
| TRA2B_HUMAN | 3.801634108 | -9.216479117 | 3.594192925 |
| PUF60_HUMAN | 1.989468646 | 1.466254652 | 3.314884359 |
| RPB2_HUMAN | 3.56347561 | -6.206428335 | 3.305719523 |
| PCBP1_HUMAN | 1.992068594 | 2.052717845 | 3.192553494 |
| SRRM2_HUMAN | 3.974038446 | -2.821225465 | 3.167279714 |
| SNW1_HUMAN | 1.989809969 | 0.46980974 | 3.052499512 |
| PRP17_HUMAN | 2.573419515 | -7.216260837 | 2.899953146 |
| RL3_HUMAN | 2.5786637 | 1.470114407 | 2.735523603 |
| TR150_HUMAN | 2.749709192 | -9.486633935 | 2.409244244 |
| BCLF1_HUMAN | 2.802239395 | -9.54041868 | 2.26048122 |
| SYF1_HUMAN | 2.453011539 | -1.520974484 | 2.259183547 |
| U2AF2_HUMAN | 0.994555861 | -8.496174702 | 2.057358601 |
| GNL3_HUMAN | 0.991971421 | 0.46977667 | 2.053387389 |
| HNRPF_HUMAN | 2.575736378 | 2.466527005 | 1.733568579 |
| AQR_HUMAN | 1.408956925 | -8.436878077 | 1.674127144 |
| XRCC6_HUMAN | 0.581108117 | -0.941562133 | 1.322646441 |
| PRP8_HUMAN | 1.132917086 | -3.823514146 | 1.143372415 |
| CLK3_HUMAN | 0.993848117 | -8.071944968 | 1.057569892 |
| OAT_HUMAN | -9.929302799 | 2.734055165 | 1.000769209 |
| CDC5L_HUMAN | 2.073375358 | -9.907942177 | 0.960108676 |
| H2A1J_HUMAN | 0.58130639 | 1.056951881 | 0.738216988 |
| ODB2_HUMAN | 0.994496198 | 2.638771496 | 0.736795312 |
| TBA1A_HUMAN | -7.574555686 | 2.466145001 | 0.735420807 |
| NOP56_HUMAN | 0.991616276 | 0.469501869 | 0.734475527 |
| HS71L_HUMAN | 0.991118205 | 0.469252722 | 0.734107857 |
| SSRP1_HUMAN | 0.99505304 | -1.522078189 | 0.544862646 |
| PPM1B_HUMAN | 0.20003204 | 1.577484201 | 0.308705987 |
| U5S1_HUMAN | 1.298784908 | -2.846251758 | 0.290901674 |
| SP16H_HUMAN | 1.315840207 | -1.519385898 | 0.060224306 |
| CASC3_HUMAN | 11.73089465 | 13.48645796 | 0 |
| WIBG_HUMAN | 11.70852352 | 10.18512321 | 0 |
| CCDC9_HUMAN | 9.914400082 | 0 | 0 |
| RL5_HUMAN | 9.168697186 | 10.22809716 | 0 |
| REN3B_HUMAN | 8.468705491 | 10.74852832 | 0 |
| BOLA2_HUMAN | 0 | 12.01531051 | 0 |
| RL13_HUMAN | 0 | 10.13647852 | 0 |
| RO52_HUMAN | 0 | 9.551362204 | 0 |
| P53_HUMAN | 0 | 9.24026662 | 0 |
| PRDX2_HUMAN | 0 | 9.229299509 | 0 |
| GSHR_HUMAN | 0 | 8.831528936 | 0 |
| C1TM_HUMAN | 0 | 8.511475946 | 0 |
| TBB2A_HUMAN | 0 | 8.063988217 | 0 |
| LAP2A_HUMAN | 0 | 7.425845306 | 0 |
| EF1A1_HUMAN (+1) | -0.22569663 | 1.249191701 | -0.068927127 |
| TBB5_HUMAN | -0.00348044 | 0.986567489 | -0.161984954 |
| IRS4_HUMAN | 0.986035447 | 2.776196272 | -0.257407108 |
| DJC10_HUMAN | -6.766065051 | 1.463811228 | -0.258872981 |
| RS8_HUMAN | 2.578589362 | 1.470068567 | -0.26084481 |
| EFTU_HUMAN | 0.410946763 | 1.692939326 | -0.261061789 |
| STK38_HUMAN | -0.003433705 | 1.056732389 | -0.261408701 |
| REST_HUMAN | -1.3198622 | 1.848156873 | -0.581625339 |
| CK084_HUMAN | -11.39580187 | 1.346444398 | -0.676241922 |
| RL4_HUMAN | 1.731826571 | 2.344806224 | -0.844616109 |
| TBA4A_HUMAN | -0.003471556 | 1.055272517 | -1.256318815 |
| ILF2_HUMAN | -8.779916297 | 1.055549917 | -1.256981736 |
| SERA_HUMAN | -0.225705018 | 0.986138055 | -1.481694665 |
| CLAP2_HUMAN | -0.58481771 | 2.051714144 | -1.827854863 |
| HNRPL_HUMAN | -8.770135276 | 1.470179053 | -1.837979591 |
| E2F7_HUMAN | -9.140190703 | 1.055919581 | -1.839883907 |
| SERPH_HUMAN | -7.683626283 | 1.467796235 | -7.683626283 |
| HDAC6_HUMAN | -0.584962501 | 1.689961089 | -7.72900887 |
| AIFM1_HUMAN | -8.129386063 | 1.054571811 | -8.129386063 |

**#Alternate EJC_replicate2**

175 proteins at 99% minimum, 2 minimum peptides, 0.0% decoy FDR;; 7282 spectra at 95% minimum, 0.00% decoy FDR

From proteins clusters, only member proteins with a non-zero value in at least one sample were retained

Normalized spectral abundance factor (NSAF) values from Scaffold were used for quantitation; a psuedo count of 0.000001 was added to each NSAF value before determining enrichment in EJC IPs as compared to FLAG-only

Proteins enriched >2-fold in any of the three EJC IPs over FLAG-only IP are shown

| | log2(NSAF) | | |
|---|---|---|---|
| Accession Number | Magoh | MLN51 | RNPS1 |
| IF4A3_HUMAN | 13.7638348 | 11.03967305 | 13.42245903 |
| RNPS1_HUMAN | 12.09199728 | 0 | 13.31783654 |
| MGN_HUMAN | 14.55895875 | 10.9479291 | 12.93590168 |
| SMD1_HUMAN | 11.36260139 | 0 | 12.86834029 |
| PININ_HUMAN | 13.14623453 | 0 | 12.80982841 |
| H2A1D_HUMAN (+6) | 12.04217235 | 0 | 12.74081375 |
| ACINU_HUMAN | 12.39159267 | 0 | 12.71647653 |
| H4_HUMAN | 0 | 0 | 12.5912652 |
| RU2A_HUMAN | 10.8482317 | 12.3128546 | 12.48507456 |
| SRSF7_HUMAN | 11.68439798 | 0 | 12.23101112 |
| SRSF1_HUMAN | 11.30320995 | 0 | 12.08921873 |
| SAP18_HUMAN | 11.32198445 | 0 | 12.02049387 |
| LC7L3_HUMAN | 11.50248377 | 0 | 12.00842862 |
| RBM39_HUMAN | 11.5997734 | 0 | 11.9935753 |
| SF3B1_HUMAN | 11.7159191 | 11.37378957 | 11.79709406 |
| SMD3_HUMAN | 0 | 0 | 11.78610623 |
| TRA2B_HUMAN | 10.6727789 | 0 | 11.69300772 |
| U5S1_HUMAN | 11.46183752 | 0 | 11.67511904 |
| RBM25_HUMAN | 11.0617087 | 0 | 11.52214012 |
| PRP19_HUMAN | 12.03479896 | 0 | 11.43827206 |
| RUXE_HUMAN | 0 | 0 | 11.4325941 |
| SRRM1_HUMAN | 11.72250872 | 0 | 11.38392025 |
| TR150_HUMAN | 11.78418496 | 0 | 11.26631806 |
| SRRM2_HUMAN | 11.4604559 | 0 | 11.23038076 |
| H13_HUMAN | 0 | 12.51929225 | 11.16835873 |
| SRSF9_HUMAN | 0 | 0 | 11.16835873 |

| Protein |  |  |  |
|---|---|---|---|
| LC7L2_HUMAN | 10.64304445 | 0 | 11.1489214 |
| PR40A_HUMAN | 10.74760559 | 0 | 11.14134082 |
| RSMB_HUMAN (+1) | 9.352175878 | 11.55252501 | 11.04937211 |
| RU2B_HUMAN | 0 | 11.32378662 | 10.94989965 |
| PRP4B_HUMAN | 9.453003062 | 0 | 10.93472289 |
| SRSF3_HUMAN | 10.48522554 | 0 | 10.92072502 |
| U520_HUMAN | 9.841595796 | 0 | 10.89565101 |
| BCLF1_HUMAN | 11.90358063 | 0 | 10.86581058 |
| SRSF6_HUMAN | 9.832177982 | 0 | 10.70017959 |
| SMD2_HUMAN | 0 | 0 | 10.65883628 |
| SNR40_HUMAN | 10.7779128 | 0 | 10.64664871 |
| SF3B6_HUMAN | 0 | 0 | 10.57572832 |
| H2B1C_HUMAN (+8) | 0 | 10.16100588 | 10.56424478 |
| SF3A2_HUMAN | 11.20683123 | 10.28019079 | 10.55803798 |
| U2AF2_HUMAN | 9.688757349 | 0 | 10.52424786 |
| PPIG_HUMAN | 9.700179593 | 0 | 10.39820926 |
| DDX41_HUMAN | 10.78422586 | 0 | 10.37634248 |
| SNW1_HUMAN | 10.89193556 | 0 | 10.35005518 |
| FIP1_HUMAN | 0 | 0 | 10.32732808 |
| CDC5L_HUMAN | 11.13359178 | 0 | 10.30924896 |
| PCBP1_HUMAN | 9.782768932 | 0 | 10.28840471 |
| SRS10_HUMAN | 10.54631572 | 0 | 10.24531501 |
| RS4X_HUMAN | 10.21916852 | 0 | 10.23983717 |
| SCML1_HUMAN | 0 | 0 | 10.17990909 |
| RBMX_HUMAN | 10.45419646 | 0 | 10.15317199 |
| PRP8_HUMAN | 8.771390425 | 0 | 10.12528431 |
| CH60_HUMAN | 11.00204143 | 0 | 10.11647393 |
| RBM22_HUMAN | 10.54361188 | 0 | 10.05012073 |
| XRCC6_HUMAN | 0 | 0 | 9.876777836 |
| H33_HUMAN | 0 | 0 | 9.86968637 |
| TRA2A_HUMAN | 0 | 0 | 9.817655232 |
| RIOK1_HUMAN | 10.27821712 | 10.79547151 | 9.807483728 |
| BUD31_HUMAN | 0 | 0 | 9.787314917 |
| CCNK_HUMAN | 9.66370041 | 0 | 9.777353958 |
| SON_HUMAN | 9.968090752 | 0 | 9.757306662 |
| GRP75_HUMAN | 10.5510353 | 0 | 9.71998768 |
| RS18_HUMAN | 0 | 0 | 9.709411269 |
| AQR_HUMAN | 9.307861087 | 0 | 9.669062209 |
| RL7_HUMAN | 0 | 0 | 9.588246152 |
| RL7A_HUMAN | 0 | 0 | 9.487297414 |
| XRCC5_HUMAN | 0 | 0 | 9.441989675 |
| ISY1_HUMAN | 9.688757349 | 0 | 9.387909618 |
| PGAM5_HUMAN | 9.08459582 | 0 | 9.367829698 |
| RS2_HUMAN | 9.064796646 | 10.52816105 | 9.348042047 |
| CSK21_HUMAN | 0 | 0 | 9.3468018 |
| SNIP1_HUMAN | 8.631140669 | 0 | 9.328495421 |
| SMU1_HUMAN | 10.42500604 | 0 | 9.277054876 |

| Protein | Value 1 | Value 2 | Value 3 |
|---|---|---|---|
| SR140_HUMAN | 9.421854445 | 10.59039992 | 9.272839701 |
| PR38A_HUMAN | 0 | 0 | 9.257529276 |
| RLA0_HUMAN | 9.535430915 | 8.831845581 | 9.234625857 |
| GPTC8_HUMAN | 0 | 0 | 9.21273955 |
| SFR19_HUMAN | 8.226074963 | 0 | 9.185494924 |
| SNUT1_HUMAN | 7.620366471 | 0 | 9.121714814 |
| PAXB1_HUMAN | 9.003855181 | 0 | 9.117435318 |
| H2A1A_HUMAN (+2) | 0 | 0 | 8.925168681 |
| RL3_HUMAN | 0 | 0 | 8.888986721 |
| HNRH1_HUMAN | 9.034001065 | 0 | 8.733388238 |
| TBA1B_HUMAN | 9.027601948 | 0 | 8.727001465 |
| PPIE_HUMAN | 0 | 0 | 8.725400344 |
| HNRPC_HUMAN | 10.00084508 | 10.4655664 | 8.701687664 |
| GRP78_HUMAN | 9.490690457 | 0 | 8.6061831 |
| PLRG1_HUMAN | 8.255972742 | 0 | 8.538848521 |
| RED_HUMAN | 8.72369507 | 0 | 8.423241996 |
| CDK12_HUMAN | 7.722602637 | 0 | 8.418738108 |
| PRP17_HUMAN | 10.08081753 | 0 | 8.3675458 |
| DDX21_HUMAN | 8.647602329 | 0 | 8.347178416 |
| RBBP6_HUMAN | 7.04537744 | 0 | 8.322694323 |
| SPF45_HUMAN | 10.61028666 | 10.07588021 | 8.312701473 |
| NKAP_HUMAN | 0 | 0 | 8.263363107 |
| RL4_HUMAN | 9.842271763 | 0 | 8.222360218 |
| SYF1_HUMAN | 8.842161743 | 0 | 8.220668531 |
| SRRT_HUMAN | 8.072427412 | 0 | 8.185767456 |
| MFAP1_HUMAN | 0 | 0 | 8.182493675 |
| OAT_HUMAN | 0 | 0 | 8.182493675 |
| CWC22_HUMAN | 8.434503133 | 0 | 8.13422094 |
| EF1A1_HUMAN (+1) | 8.409348503 | 0 | 8.109099329 |
| SRS11_HUMAN | 0 | 0 | 8.04220656 |
| HNRPM_HUMAN | 9.33230514 | 0 | 8.034358798 |
| DHX8_HUMAN | 0 | 0 | 8.03039126 |
| ZC3HE_HUMAN | 9.734997453 | 0 | 8.022589749 |
| SPTN1_HUMAN | 8.574139362 | 6.877866923 | 8.011674533 |
| TAB1_HUMAN | 8.867618521 | 9.162265434 | 7.984019633 |
| MATR3_HUMAN | 0 | 0 | 7.820753373 |
| CRNL1_HUMAN | 8.119096751 | 0 | 7.819029455 |
| CHERP_HUMAN | 8.008204497 | 0 | 7.708256223 |
| CDK13_HUMAN | 0 | 0 | 7.401903472 |
| MYH9_HUMAN | 9.229515825 | 9.372473325 | 7.34978987 |
| TBB4B_HUMAN | 9.046878442 | 8.343807528 | 7.168020189 |
| ZCH18_HUMAN | 0 | 0 | 7.070067262 |
| NKTR_HUMAN | 0 | 0 | 7.037821465 |
| IPO8_HUMAN | 8.564416278 | 8.124017889 | 6.949184722 |
| CAMP3_HUMAN | 0 | 0 | 6.683134985 |
| MAP1B_HUMAN | 0 | 0 | 6.29000041 |
| SPTB2_HUMAN | 6.649256178 | 0 | 5.775182264 |

| Protein |  |  |  |
|---|---|---|---|
| FLNA_HUMAN | 7.218393711 | 0 | 5.615210284 |
| IF4B_HUMAN | 2.319331128 | 3.873106719 | 1.625925187 |
| RBM8A_HUMAN | 11.39943816 | 0 | 0 |
| CASC3_HUMAN | 11.19241612 | 12.26488238 | 0 |
| CCDC9_HUMAN | 10.20554891 | 0 | 0 |
| COR1C_HUMAN | 9.691795853 | 10.24946904 | 0 |
| REN3B_HUMAN | 9.664678382 | 0 | 0 |
| PYM1_HUMAN | 9.586314394 | 0 | 0 |
| HNRPK_HUMAN | 8.989791843 | 0 | 0 |
| HNRPR_HUMAN | 8.953701754 | 0 | 0 |
| SRSF4_HUMAN | 8.896483781 | 0 | 0 |
| SLU7_HUMAN | 8.65065687 | 0 | 0 |
| FL2D_HUMAN | 8.631140669 | 0 | 0 |
| HNRPF_HUMAN | 8.563692024 | 0 | 0 |
| OTUD4_HUMAN | 7.727103604 | 0 | 0 |
| LIMA1_HUMAN | 7.695854658 | 8.572851782 | 0 |
| TBA1C_HUMAN | 7.454504938 | 0 | 0 |
| K2C8_HUMAN | 7.349878311 | 0 | 0 |
| TMOD3_HUMAN | 0 | 11.26315181 | 0 |
| PRDX1_HUMAN | 0 | 11.0860035 | 0 |

**Table S2.** Summary of all described RIPiT-Seq and RNA-Seq libraries.

| *All numbers represent unique counts |
| **Tables A-D show counts in sequential order of processing |

**Table A.** Multi-mapping and PCR duplicate read removal

| Library | TopHat Input Total | Mapped | (%) | Multi-Mapped | (%) | Input Reads | Multi-Mappers | (%) | PCR Duplicates | (%) |
|---|---|---|---|---|---|---|---|---|---|---|
| MAGOH-EIF4A3 | 12289132 | 11788001 | 95.92 | 919103 | 7.8 | 11788001 | 919103 | 7.8 | 38395 | 0.33 |
| CASC3-EIF4A3 | 7918724 | 7552965 | 95.38 | 501748 | 6.64 | 7552965 | 501748 | 6.64 | 14235 | 0.19 |
| RNPS1-EIF4A3 | 9353236 | 8964991 | 95.85 | 921854 | 10.28 | 8964991 | 921854 | 10.28 | 16528 | 0.18 |
| MAGOH-CASC3 | 6792045 | 6423758 | 94.58 | 408153 | 6.35 | 6423758 | 408153 | 6.35 | 31314 | 0.49 |
| MAGOH-RNPS1 | 11996168 | 11169040 | 93.11 | 617910 | 5.53 | 11169040 | 617910 | 5.53 | 35527 | 0.32 |
| CASC3-EIF4A3-XL-A | 5718072 | 5309270 | 92.85 | 370143 | 6.97 | 5309270 | 370143 | 6.97 | 10068 | 0.19 |
| CASC3-EIF4A3-XL-B | 21898851 | 20055735 | 91.58 | 3554071 | 17.72 | 20055735 | 3554071 | 17.72 | 2013494 | 10.04 |
| RNPS1-EIF4A3-XL-A | 8370005 | 7838852 | 93.65 | 651639 | 8.31 | 7838852 | 651639 | 8.31 | 23133 | 0.3 |
| RNPS1-EIF4A3-XL-B | 24859586 | 23434622 | 94.27 | 4666719 | 19.91 | 23434622 | 4666719 | 19.91 | 715018 | 3.05 |
| CASC3-EIF4A3-XL-A (+CHX) | 15795694 | 15084794 | 95.5 | 684318 | 4.54 | 15084794 | 684318 | 4.54 | 182916 | 1.21 |
| CASC3-EIF4A3-XL-B (+CHX) | 13937900 | 13413713 | 96.24 | 523380 | 3.9 | 13413713 | 523380 | 3.9 | 75455 | 0.56 |
| RNPS1-EIF4A3-XL-A (+CHX) | 27687914 | 26580410 | 96 | 2258708 | 8.5 | 26580410 | 2258708 | 8.5 | 187891 | 0.71 |
| RNPS1-EIF4A3-XL-B (+CHX) | 2532100 | 2274851 | 89.84 | 95329 | 4.19 | 2274851 | 95329 | 4.19 | 26077 | 1.15 |
| RNA-Seq-A | 18740947 | 12180518 | 64.99 | 4873419 | 40.01 | 12180518 | 4873419 | 40.01 | 58947 | 0.48 |
| RNA-Seq-B | 19476485 | 12592943 | 64.66 | 4811993 | 38.21 | 12592943 | 4811993 | 38.21 | 46021 | 0.37 |

**Table B.** Removal of miscellaneous RNAs

| Library | Mitochondrial | (%) | miRNA | (%) | (%) | rRNA (+ChrUn) | (%) | scaRNA | (%) | snoRNA | (%) | snRNA | (%) | tRNA | (%) | Final Count |
|---|---|---|---|---|---|---|---|---|---|---|---|---|---|---|---|---|
| MAGOH-EIF4A3 | 144700 | 1.23 | 1384 | 0.01 | 0 | 3524726 | 29.9 | 396 | 0 | 20182 | 0.17 | 677876 | 5.75 | 2223 | 0.02 | 6459016 |
| CASC3-EIF4A3 | 124075 | 1.64 | 787 | 0.01 | 0 | 4946036 | 65.48 | 102 | 0 | 4582 | 0.06 | 462939 | 6.13 | 3095 | 0.04 | 1495366 |
| RNPS1-EIF4A3 | 108423 | 1.21 | 879 | 0.01 | 0 | 3215430 | 35.87 | 193 | 0 | 7039 | 0.08 | 1275182 | 14.22 | 3955 | 0.04 | 3415508 |
| MAGOH-CASC3 | 35630 | 0.55 | 936 | 0.01 | 0 | 3663566 | 57.03 | 105 | 0 | 3038 | 0.05 | 163479 | 2.54 | 4305 | 0.07 | 2113232 |
| MAGOH-RNPS1 | 60106 | 0.54 | 1511 | 0.01 | 0 | 5766613 | 51.63 | 133 | 0 | 4938 | 0.04 | 385709 | 3.45 | 4516 | 0.04 | 4292077 |
| CASC3-EIF4A3-XL-A | 22902 | 0.43 | 1264 | 0.02 | 0 | 3382695 | 63.71 | 322 | 0.01 | 7479 | 0.14 | 140262 | 2.64 | 3892 | 0.07 | 1370243 |
| CASC3-EIF4A3-XL-B | 98373 | 0.49 | 3261 | 0.02 | 0 | 11708237 | 58.38 | 1866 | 0.01 | 6384 | 0.03 | 1297149 | 6.47 | 10183 | 0.05 | 1362717 |
| RNPS1-EIF4A3-XL-A | 28638 | 0.37 | 1455 | 0.02 | 0 | 3253080 | 41.5 | 896 | 0.01 | 16125 | 0.21 | 506409 | 6.46 | 4484 | 0.06 | 3352993 |
| RNPS1-EIF4A3-XL-B | 50751 | 0.22 | 6254 | 0.03 | 0 | 9460234 | 40.37 | 3923 | 0.02 | 13060 | 0.06 | 2295813 | 9.8 | 8177 | 0.03 | 6214673 |
| CASC3-EIF4A3-XL-A (+CHX) | 47903 | 0.32 | 14652 | 0.1 | 0 | 4289703 | 28.44 | 334 | 0 | 7472 | 0.05 | 76118 | 0.5 | 8924 | 0.06 | 9772454 |
| CASC3-EIF4A3-XL-B (+CHX) | 60192 | 0.45 | 6725 | 0.05 | 0 | 4702167 | 35.05 | 243 | 0 | 4686 | 0.03 | 71546 | 0.53 | 6357 | 0.05 | 7962962 |
| RNPS1-EIF4A3-XL-A (+CHX) | 132946 | 0.5 | 2653 | 0.01 | 0 | 12002630 | 45.16 | 1679 | 0.01 | 36309 | 0.14 | 1267031 | 4.77 | 21014 | 0.08 | 10669549 |
| RNPS1-EIF4A3-XL-B (+CHX) | 3478 | 0.15 | 110 | 0 | 0 | 1526002 | 67.08 | 49 | 0 | 235 | 0.01 | 9165 | 0.4 | 93 | 0 | 614313 |
| RNA-Seq-A | 507860 | 4.17 | 11319 | 0.09 | 0 | 693781 | 5.7 | 5463 | 0.04 | 183690 | 1.51 | 481999 | 3.96 | 1883466 | 15.46 | 3480574 |
| RNA-Seq-B | 535924 | 4.26 | 13299 | 0.11 | 0 | 506876 | 4.03 | 6870 | 0.05 | 298611 | 2.37 | 673413 | 5.35 | 1945855 | 15.45 | 3754081 |

**Table C.** Genomic read distribution (51% overlap)

| Library | Exonic | (%) | RPKM | Intronic | (%) | RPKM | Intergenic | (%) | 1 Exon Genes | RPKM | 2+ Exon Genes | RPKM |
|---|---|---|---|---|---|---|---|---|---|---|---|---|
| MAGOH-EIF4A3 | 4384496.00 | 67.88 | 7.40 | 1797361.00 | 27.83 | 0.19 | 277159.00 | 4.29 | 30122.00 | 1.70 | 4473662.00 | 7.78 |
| CASC3-EIF4A3 | 906277.00 | 60.61 | 6.61 | 504613.00 | 33.75 | 0.23 | 84476.00 | 5.65 | 11554.00 | 2.82 | 924113.00 | 6.94 |
| RNPS1-EIF4A3 | 2162371.00 | 63.31 | 6.90 | 1098461.00 | 32.16 | 0.22 | 154676.00 | 4.53 | 23011.00 | 2.46 | 2209957.00 | 7.27 |
| MAGOH-CASC3 | 1476411.00 | 69.87 | 7.61 | 531464.00 | 25.15 | 0.17 | 105357.00 | 4.99 | 15495.00 | 2.68 | 1505597.00 | 8.00 |
| MAGOH-RNPS1 | 3046675.00 | 70.98 | 7.74 | 1026440.00 | 23.91 | 0.16 | 218962.00 | 5.10 | 30176.00 | 2.57 | 3105684.00 | 8.13 |
| CASC3-EIF4A3-XL-A | 914747.00 | 66.76 | 7.28 | 379830.00 | 27.72 | 0.19 | 75666.00 | 5.52 | 7480.00 | 1.99 | 936445.00 | 7.68 |
| CASC3-EIF4A3-XL-B | 1193285.00 | 87.57 | 9.54 | 692445.00 | 50.81 | 0.34 | 523013.00 | 38.38 | 16125.00 | 4.32 | 1231702.00 | 10.15 |
| RNPS1-EIF4A3-XL-A | 2274827.00 | 67.84 | 7.39 | 908998.00 | 27.11 | 0.18 | 169168.00 | 5.05 | 15559.00 | 1.70 | 2336427.00 | 7.83 |
| RNPS1-EIF4A3-XL-B | 7357943.00 | 118.40 | 12.90 | 3281206.00 | 52.80 | 0.36 | 4424476.00 | 71.19 | 53784.00 | 3.16 | 7573547.00 | 13.69 |
| CASC3-EIF4A3-XL-A (+CHX) | 7860415.00 | 80.43 | 8.77 | 1741791.00 | 17.82 | 0.12 | 170248.00 | 1.74 | 47235.00 | 1.77 | 8010464.00 | 9.21 |
| CASC3-EIF4A3-XL-B (+CHX) | 6528217.00 | 81.98 | 8.94 | 1331364.00 | 16.72 | 0.11 | 103381.00 | 1.30 | 26773.00 | 1.23 | 6674856.00 | 9.42 |
| RNPS1-EIF4A3-XL-A (+CHX) | 7057878.00 | 66.15 | 7.21 | 3096590.00 | 29.02 | 0.20 | 515081.00 | 4.83 | 46214.00 | 1.58 | 7205508.00 | 7.59 |
| RNPS1-EIF4A3-XL-B (+CHX) | 454507.00 | 73.99 | 8.06 | 117397.00 | 19.11 | 0.13 | 42409.00 | 6.90 | 4588.00 | 2.73 | 463153.00 | 8.47 |
| RNA-Seq-A | 1059522.00 | 30.44 | 3.32 | 1761478.00 | 50.61 | 0.34 | 659574.00 | 18.95 | 81172.00 | 8.52 | 1064409.00 | 3.44 |
| RNA-Seq-B | 1080657.00 | 28.79 | 3.14 | 1785669.00 | 47.57 | 0.32 | 887755.00 | 23.65 | 110593.00 | 10.76 | 1072926.00 | 3.21 |

**Table D.** Canonical/Non-canonical read classification

| Library | Canonical | (%ofExonic) | Non-Canonical | (%ofExonic) | Uncounted | (%ofExonic) |
|---|---|---|---|---|---|---|
| MAGOH-EIF4A3 | 1546096.00 | 35.26 | 2722620.00 | 62.10 | 115780.00 | 2.64 |
| CASC3-EIF4A3 | 289135.00 | 31.90 | 579485.00 | 63.94 | 37657.00 | 4.16 |
| RNPS1-EIF4A3 | 811486.00 | 37.53 | 1267130.00 | 58.60 | 83755.00 | 3.87 |
| MAGOH-CASC3 | 476638.00 | 32.28 | 951547.00 | 64.45 | 48226.00 | 3.27 |
| MAGOH-RNPS1 | 898705.00 | 29.50 | 2006039.00 | 65.84 | 141931.00 | 4.66 |
| CASC3-EIF4A3-XL-A | 326669.00 | 35.71 | 530307.00 | 57.97 | 57771.00 | 6.32 |
| CASC3-EIF4A3-XL-B | 587174.00 | 49.21 | 543975.00 | 45.59 | 62136.00 | 5.21 |
| RNPS1-EIF4A3-XL-A | 723741.00 | 31.82 | 1407464.00 | 61.87 | 143622.00 | 6.31 |
| RNPS1-EIF4A3-XL-B | 2555021.00 | 34.72 | 4178245.00 | 56.79 | 624677.00 | 8.49 |
| CASC3-EIF4A3-XL-A (+CHX) | 2386087.00 | 30.36 | 4692151.00 | 59.69 | 782177.00 | 9.95 |
| CASC3-EIF4A3-XL-B (+CHX) | 3145982.00 | 48.19 | 2672452.00 | 40.94 | 709783.00 | 10.87 |
| RNPS1-EIF4A3-XL-A (+CHX) | 1593969.00 | 22.58 | 4853509.00 | 68.77 | 610400.00 | 8.65 |
| RNPS1-EIF4A3-XL-B (+CHX) | 46447.00 | 10.22 | 358889.00 | 78.96 | 49171.00 | 10.82 |

**Table S3.** List of reagents (siRNA sequences, qRT-PCR primers, antibodies, plasmids)

### siRNAs

| Gene | siRNA sequence | Reference |
|---|---|---|
| *AllStars Negative Control* | *Qiagen proprietary sequence* | |
| *EIF4A3_187* | CGA GCA AUC AAG CAG AUC AUU | This Study |
| *CASC3_534* | CCA GCA UAC AUA CCU CGG AUU | This Study |
| *RNPS1_2* | GCA UCC AGC CGC UCA GGA AUU | This Study |

### qRT-PCR primers

| Gene | Primer (Forward) | Primer (Reverse) | Reference |
|---|---|---|---|
| *GADD45A* | GAGCTCCTGCTCTTGGAGAC | GCAGGATCCTTCCATTGAGA | This Study |
| *UPP1* | CCAGCCTTGTTTGGAGATGT | ACATGGCATAGCGGTCAGTT | This Study |
| *ARC* | CTGAGATGCTGGAGCACGTA | GCCTTGATGGACTTCTTCCA | This Study |
| *DNAJB2* | TGGCATCCTACTACGAGATCC | GTTTTTGTCTGGGTGCCACT | This Study |
| *eIF4A2* | TGTGGAGACGGTGACAGATTC | TTCCTGCTTTACCCACCAGTAC | This Study |
| *TMEM33* | AGGGTCAAGTCGTGTTCTGATC | ACCAACTGCTGCTATCGACTC | This Study |
| *SF3B1* | CACGGCAATGTGGCTTTCTC | ACTCACCAAGAAGAGGCAGAAC | This Study |
| *CAPN7* | AATTTCCCCAGAGCGTCTTG | TTCGTGCCTTTGTCTCCATC | This Study |
| *TPT1* | CAACGGGAAGGCGAGCTCTC | GGAAGGAGCGGCAAAGTTTAC | This Study |
| *C1orf37* | TCATCAGCCATGGTCAGTAGG | TGCAGGTGATGGTTCATGAC | This Study |
| *SRSF2* | TTGCTGCTCGAATCTCCAAG | ACTTCTGCTGCCATCACAAC | Lareau et al., 2007 |
| *SRSF4* | CCTCTTAAGAAAATGCTGCGGTCTC | ATCAGCCAAATCAGTTAAAATCTGC | Lareau et al., 2007 |
| *SRSF6* | GGATCTGAAGAACGGTCTGTTATGT | TCACTCGTCTTTTGGTTCCCATTAG | Lareau et al., 2007 |
| *RPL3* | TGGTGGAAAAGGTCCTTAG | TCAGGAGCAGAGCAGAGTTG | This Study |
| *RPL12* | CTGGGCCTTAGCTTCTTCAC | AAGTGGCACCGACTTCACCT | This Study |
| *ACTB* | CGCGAGAAGATGACCCAGAT | TCACCGGAGTCCATCACGAT | This Study |

### Antibodies

| Protein | Source | Catalog# |
|---|---|---|
| CASC3 | Bethyl Laboratories | A302-472 |
| RNPS1 | Sigma Aldrich | HPA044014 |
| RNPS1 | Akila Mayeda | |
| EIF4A3 | Bethyl Laboratories | A302-980A |
| HNRNPA1 | Santa Cruz Biotechnology | SC-32301 |
| ACIN1 | Bethyl Laboratories | A300-999A |
| SAP18 | Santa Cruz Biotechnology | SC-25377 |
| RBM8A | Sigma Aldrich | Y1253 (4C4) |
| MAGOH | Santa Cruz Biotechnology | SC-271365 |
| SRSF1 | Santa Cruz Biotechnology | SC-33652 |
| FLAG | Sigma Aldrich | F1804 |

**Reagents**

| Reagent | Source | Catalog# |
|---|---|---|
| Protein A Dynabeads | Life Technologies | 10002D |
| Protein G Dynabeads | Life Technologies | 10003D |
| FLAG-Agarose Beads | Sigma Aldrich | A2220 |
| FLAG peptide | Sigma Aldrich | F3290 |
| MWCO 7,000 Da Dialysis Column | Pierce | 88242 |
| 4X Lammelli SDS load buffer | Bio-Rad | 161-0737 |
| 4%–15% Mini-PROTEAN TGX | Bio-Rad | 4561086 |
| Trypsin Gold | Promega | V5280 |
| ProteaseMAX™ Surfactant | Promega | V2071 |
| TransIT-X2 | Mirus | MIR 6003 |
| RNAiMAX | Thermo Fisher | 13778030 |
| JetPrime | PolyPlus | 114-01 |
| oligo-(dT)15 Primer | Promega | C1101 |
| Superscript III | Invitrogen | 18080044 |
| Ribonuclease H | Promega | M4281 |
| 2X SYBR Green Master Mix | Applied Biosystems | 4309155 |
| RNase I | Promega | M4261 |
| Ribo-Zero rRNA Removal Kit (H/M/R) | Illumina | MRZH116 |

**Plasmids**

| Plasmid | Reference |
|---|---|
| pcβwtβ | Lykke-Andersen, 2000 |
| pcβ39β | Lykke-Andersen, 2000 |
| pcTET2-βwtβ | Singh et al., 2007 |
| pcTET2-β39β | Singh et al., 2007 |
| pcβwtGAP3UAC | Lykke-Andersen, 2000 |
| pcDNA3 | Singh et al., 2012 |
| pcDNA3 YFP | This study |
| pcDNA3 FLAG | Singh et al., 2012 |
| pcDNA3 FLAG-EIF4AIII | Singh et al., 2012 |
| pcDNA3 FLAG-EIF4AIII YRAA | This study |
| pcDNA3 FLAG-MAGOH | Singh et al., 2012 |
| pcDNA3 FLAG-RNPS1 | This study |
| pcDNA3 FLAG-CASC3 | This study |
| pcDNA3 FLAG-HNRNPA1 | This study |
| pcDNA5 FLAG-MAGOH | Singh et al., 2012 |
| pcDNA5 FLAG-RNPS1 | This study |
| pcDNA5 FLAG-CASC3 | This study |

# EXPERIMENTAL PROCEDURES

## Plasmids, Cell lines and cell culture

Plasmids expressing EJC factors and hnRNPA1 were generated by cloning their cDNAs into derivatives of pcDNA3.1 (for transient expression) or pcDNA5-FLAG-FRT/TO (for stable expression) as described previously (Singh et al., 2012). Plasmids expressing β-globin reporters have been described elsewhere (Lykke-Andersen, 2000; Singh, 2007). Human cell lines (HEK293 Flp-In TRex, HeLa CCL2, HeLa Tet-off) were cultured in Dulbecco's modified eagle medium (DMEM) supplemented with 10 % fetal bovine serum (FBS) and 1 % penicillin-streptomycin. Stable cell lines expressing tetracycline-inducible FLAG-tagged proteins were created using HEK293 Flp-In TRex cells as described previously (Singh et al., 2012).

## Endogenous Immunoprecipitations

HEK293, HeLa and P19 cells were lysed and sonicated in hypotonic lysis buffer (HLB) [20 mM Tris-HCl pH 7.5, 15 mM NaCl, 10 mM EDTA, 0.5 % NP-40, 0.1 % Triton X-100, 1 x Sigma protease inhibitor cocktail, 1 mM PMSF]. Lysates were sonicated, increased to 150 mM NaCl and treated with RNase A for five minutes. Complexes were then captured on Protein A/G Dynabeads (Life Technologies) conjugated to IgG, α-EIF4A3, α-CASC3 or α-RNPS1 antibodies for 2 hr nutating at 4 °C. Complexes were washed in isotonic wash buffer (IsoWB) [20 mM Tris-HCl pH 7.5, 150 mM NaCl, 0.1 % NP-40] and eluted in clear sample buffer [100 mM Tris-HCl pH 6.8, 4 % SDS, 10 mM EDTA, 100 mM DTT]. Cortical neurons were isolated from FBV wild-type male mice in HBSS buffer. Once isolated, lysate preparation was carried out as stated above.

## FLAG Immunoprecipitations

Stable HEK293 cells expressing FLAG-tagged EJC protein were lysed in HLB. Lysates were sonicated, increased to 150 mM NaCl and treated with RNase A. Complexes were then captured on FLAG-beads, washed with IsoWB and FLAG-peptide eluted. IPs were carried out as noted above.

## Mass Spectrometry

*FLAG Immunoprecipitation*

Stably expressed FLAG-MAGOH, FLAG-CASC3, FLAG-RNPS1, or FLAG-peptide were IPed and RNase A digested at 4 °C for 2 hr in lysis buffer supplemented with 1 µg/ml FLAG peptide to improve IP specificity. The FLAG affinity elution was completely dried by vacuum evaporation, re-suspended in 100 µl of water and dialyzed (dialysis buffer: 10 mM Tris-HCl pH7.5, 75 mM NaCl, 0.01 % Triton X-100) for ~6 hr at 4 °C in a MWCO 7,000 Da mini dialysis column (Pierce). The dialyzed sample (90-100 µl) was again completely dried by vacuum evaporation and re-suspended in 20 µl of 0.1 % SDS and 10 mM DTT (prepared from a fresh stock solution). The sample was heated at 95 °C for 5 min and cooled to room temperature. The reduced thiol groups were alkylated by incubating with 0.8 µl of freshly prepared 1M iodoacetamide at room temperature for 45 min in dark. The resulting samples were mixed with 5 µl of 4 x Lammelli SDS load buffer (Bio-Rad) and loaded on 4 %–15 % Mini-PROTEAN TGX gel (Bio-Rad). The samples were migrated until the dye-front had run approximately 1 cm into the gel from the bottom of well. The gel was washed three times with ~200 ml HPLC grade water for 5 min each. The gel piece containing protein was excised and processed for in gel digestion of proteins.

*In-Gel Digestion*

The gel slices were cut into 1×5×1mm$^3$ dimension and transferred into a 1.5 mL microcentrifuge tube. Samples were washed with water and dehydrated in acetonitrile :50 mM $NH_4HCO_3$ (1:1 v/v) for 5 minutes and 100% acetonitrile for 30 seconds. Following drying with a vacuum concentrator (Savant), gel slices were rehydrated in 25 mM dithiothreitol in 50 mM $NH_4HCO_3$ and incubated for 20 min at 56 °C. With the removal of the supernatant, 55 mM iodoacetamide in 50 mM $NH_4HCO_3$ was added and the samples were incubated in the dark for 20 min at room temperature. Gel slices were washed and dehydrated as described before. Following removal of the liquid by vacuum concentrator, the sample was rehydrated in 12 ng/µL Trypsin Gold (Promega) in 0.01 % ProteaseMAX$^{TM}$ Surfactant (Promega): 50 mM $NH_4HCO_3$ and incubated at 50 °C for one hr. Condensate was collected by centrifuging at 14,000 × g for 10 seconds and the solution was transferred to a new tube. Trifluoroacetic acid was added to a final concentration of 0.5 % to inactivate trypsin and the solution was dried by vacuum concentrator.

*LC/MS/MS*

The LC-MS/MS tryptic digested peptides were dissolved in 0.1 % formic acid (v/v) and loaded at 15 µL/min for 10 min using a nanoACQUITY C18 Trap column (20×0.18 mm i.d., 5 µm 100Å C18, Waters, 186007238). Peptides were separated using an EASY-Spray LC Column(150×0.075 mm i.d., 3 µm 100Å C18, ThermoFisher, ES800) with a gradient using solvent A (0.1 % formic acid in water) and solvent B (0.1 % formic acid in acetonitrile) at a 0.5 µL/min flow rate. The gradient started with 1 % B for 5 min, followed by 85 min with a linear increase to 35 % B. The gradient was increased to 85 % B in 5 min followed by decreasing back to 1 % B in 5 min as the final wash step. A nanoACQUITY UPLC (Waters) was coupled to a Velos Pro Dual-Pressure Linear Ion Trap mass spectrometer (Thermo Scientific) for data acquisition. A data-dependent acquisition routine was used for a mass spectrum from m/z 300 to 2000 and followed by ten tandem mass spectrometry scans.

*Data Analysis*

Raw data files were processed using Proteome Discoverer (ThermoFisher). The data were searched against the human Swiss-Port index (09/01/15) using Sequest HT with precursor mass tolerances of 1.5 Da and fragment mass tolerances of 0.8 Da. Maximum missed cleavage sites of full tryptic digestion was two and dynamic modifications of acetylation (N-terminus), carbamidomethylation(cysteine), propionamidation(cysteine) and oxidation(methionine) were considered. The processed mass spectrometry data was analyzed using the Scaffold software (4.4.5). Quantifications from replicate 1 are presented in main figures and from replicate 2 in supplementary figures. For the analysis presented in Figure 1 and S1, spectral quantification was done using the normalized weighted spectral counts with minimum peptide identification threshold at 95 % and protein identification threshold of 95 % with 1 minimum peptide. For analysis presented in Figure 2 and S2, spectral quantification was done using the Normalized Spectral Abundance Factor (NSAF) (Zhang et al., 2010). A total of 341 (replicate 1) or 259 (replicate 2) proteins were identified at 99 % threshold each with a minimum of 2 peptides. When Scaffold grouped together similar proteins into clusters, only those proteins were retained from a cluster that had non-zero NSAF value in at least one of the four samples. There are several proteins that are detected only in one or the other alternate EJC, and thus had zero values in samples where they were undetected. Also, many EJC specific proteins had zero values in FLAG-only control IPs. A pseudocount of 0.00001 was added to all NSAF values to prevent loss of such proteins when divided by zero. Fold-enrichment over FLAG-only control was calculated for each of the three EJC samples as follows:

Fold-enrichment:
$$\log_2[NSAF(FLAG\text{-}EJC+0.00001)] / [NSAF(FLAG\text{-}only+0.000001)]$$

The comparison of fold-enrichment values for all 341 proteins in the EJC core or alternate EJCs was carried out in R package ggplot2 (via scatter plots). Proteins that were >10-fold enriched in one of the three EJC IPs were included in the heatmap (using R package gplots). In this analysis, proteins that most likely were contaminants (e.g. cytoskeleton proteins, histones, metabolic enzymes) were not included in the heatmap but are shown in the >10-fold enrichment table.

**Glycerol Gradient Fractionation**

Five 15-cm plates with ~90 % confluent HEK293 cells expressing FLAG-tagged MAGOH, CASC3, RNPS1 proteins, or FLAG-peptide as a control, were cultured and induced as above. Cell lysis, FLAG-immunoprecipitation, RNase A digestion and FLAG-elution steps were also carried out as described above. FLAG-IP elution was layered onto pre-cooled continuous 10-30 % glycerol gradients prepared in 11 ml Beckman centrifuge tubes. Gradients were run at 32,000 rpm for 16 hours at 4 °C. Gradients were fractionated by-hand into 500 µl fractions. Proteins were precipitated using TCA and resuspended in 15 µl of 1x SDS loading buffer for analysis on 12 % SDS-PAGEs.

**siRNA-mediated Knockdowns**

*HEK293 siRNA knockdowns*

HEK293 cells were seeded into 24-well plates. Cells were transfected 4 hours later following the Mirus TransIT-X2 procedure with 55-60 pmols of control, EIF4A3, CASC3, RNPS1, or UPF1 siRNA. Cells were harvested 48 hr later. Knockdown was checked by Western blot and subsequently purified RNA was used for qRT-PCR analysis. For 96 hr knockdowns, HEK293 cells were transfected as above and were incubated for 24 hr. Cells were transfected again at roughly 70-80 % confluence following the Mirus TransIT-X2 procedure using 50-60 pmol of control, EIF4A3, CASC3, RNPS1, or UPF1 siRNA. Cells were harvested ~48 hr (~96 total hours after first transfection) later. Knockdown was checked by Western blot and subsequently purified RNA was used for qRT-PCR analysis.

*HeLa Tet-off siRNA knockdowns and reporter RNA expression*

HeLa Tet-off cells in 12-well plates (1.2 x $10^5$ cells/well) were reverse transfected with 15 pmol of control, UPF1, CASC3 and RNPS1 siRNA using RNAiMAX following the manufacturer's protocol. After 24 hr, cells were co-transfected with plasmids (100 ng of pcTET2-βwtβ or pcTET2-β39β, 30 ng of pcβwtGAP3UAC and 70 ng of pezYFP or carrier DNA) and a second dose of 15 pmol of siRNA using JetPrime (PolyPlus). Tetracycline (50 ng/ml) was included at the time of transfection to repress reporter RNA expression, and was removed 20 hr later to induce reporter RNA expression for 6-8 hr. Cells were harvested in clear sample buffer for western blot analysis and TRIzol extraction of RNA, which was analyzed by Northern Blotting.

**Overexpression of alternate EJC factors**

*HEK293 FLAG-EJC overexpression*

HEK293 cells expressing FLAG-tagged CASC3, RNPS1 proteins, or FLAG-peptide as a control, were seeded into 6-well plates. Cells were induced with 625 ng/ml of Doxycycline for 48 hr. Overexpression was checked by Western blot and subsequently purified RNA was used for qRT-PCR analysis.

For IPs following alternate EJC protein overexpression, HEK293 cells stably expressing FLAG-tagged CASC3 or RNPS1, or FLAG-peptide as a control, were seeded in 10 cm plates and induced with 625 ng/ml of Doxycycline for 48 hr. Endogenous RBM8A or EIF4A3 IPs were carried out as described above.

**HEK293 EIF4A3 or EIF4A3 YRAA Rescue**

HEK293 cells were seeded into 24-well plates. Cells were transfected 4 hr later following the Mirus TransIT-X2 procedure with 60 pmol of control, or EIF4A3 siRNA, and 300 ng of pcDNA3 empty, FLAG-EIF4A3 wt or FLAG-EIF4A3 YRAA and 200 ng of pcDNA3 vector. Cells were harvested ~48 hr later. Knockdown and EIF4AIII expression was checked by Western blot and subsequently purified RNA was used for qRT-PCR analysis.

HEK293 cells were seeded into 6-well plates. Cells were transfected four hours later following the Mirus TransIT-X2 protocol with 300 ng of pcDNA3 empty, FLAG-EIF4A3 wt or FLAG-EIF4A3 YRAA and 200 ng of pcDNA3 vector. Cells were harvested in 1x PBS 48 hr later and washed for subsequent FLAG IP as previously described.

**β-globin NMD assays**

The pulse-chase experiments were performed in HeLa Tet-off cells as described previously (Singh et al., 2007). Briefly, cells growing in 12-well plates were transfected with 10 ng of pcβwtGAP3UAC, 200 ng of pcTET2-β39β and 250 ng of pezFLAG-CASC3 (or pezFLAG empty vector) in the presence of 50 ng/nl tetracycline. 36 hr later, expression of β39 mRNA was induced for 6 hr by removing Tet, and then supressed again via addition of 1 μg/ml Tet. Time points were collected starting ~30 min after addition of Tet for 0 hr time point. Total RNA was extracted using TRIzol and one-half of the RNA sample was analyzed by Northern blots. The autoradiogram signal was scanned using Fuji FLA imager, and quantified using ImageQuant software.

For steady-state assays, HeLa CCL2 cells growing in 12-well plates were transfected with 100 ng of pcβwtGAP3UAC, 100 ng of either pcβwtβ or pcβ39β, and 150 ng of pezFLAG plasmids (expressing RNPS1 or CASC3, or empty vector as a control). Cells were harvested 48 hr post-transfections and total RNA extracted was analyzed by Northern blots.

**qRT-PCR**

RNA was isolated from cells using Trizol, DNase treated, purified with Phenol:Chloroform:Isoamyl alcohol (25:24:1, pH 4.5) and resuspended in RNase-free water. RNA was reverse transcribed using oligo-dT (Promega) and Superscript III (Invitrogen). After reverse transcription of RNA the samples were treated with RNase H (Promega) for 30 min at 37 °C. The samples were then diluted to 5 ng/μl before proceeding to qPCR setup. For each qPCR 25 ng of cDNA was mixed with 7.5 μl of 2 x SYBR Green Master Mix (ABS), 0.6 μl of a 10 mM forward and reverse primer mix (defrosted once), and quantity sufficient water for a 15 μl reaction volume. The qPCRs were performed on (Applied Biosystems) in triplicates (technical). Parallel with NMD targets, β-actin or TATA-binding protein (TBP) were used as internal controls. Fold-change calculations were performed by delta-delta Ct method. Fold-change from at least three biological replicates were used to determine the standard error of means. The p-values were calculated using student's t-test.

**Cytoplasmic FLAG-IPs of alternate EJCs**

One 15-cm plate of HEK293 cells expressing FLAG-tagged MAGOH, CASC3, RNPS1 proteins, or FLAG-peptide as a control, were cultured and induced as above. Cells were lysed in 2 ml RSB-150 buffer [10 mM Tris-HCl pH 7.5, 150 mM NaCl, 5 mM $MgCl_2$, 10 μg/ml Aprotinin, 10 μM Leupeptin, 1 μM Pepstatin, 1 mM PMSF] supplemented with 0.05 % digitonin and 1 mM DTT. Lysates were passed through a 25 gauge needle 5 times before centrifugation at 3,000 x g for 1 min at 4 °C. Supernatant was removed and centrifuged at 21,000 x g for 10 min at 4 °C, and passed through a 0.45 μm filter (Cytoplasmic Fraction). The first pellet was resuspended in RSB-150 and pelleted again at 3,000 x g for 1 min at 4 °C to remove residual cytoplasmic contaminants. The nuclei were then resuspended in 2 ml RSBT (components) and sonicated as previously described, before being centrifuged and purified as done for the cytoplasmic fraction.

Approximately 2 ml of each fraction were collected and used for FLAG-IP as described above.

**RIPiT-Seq data analysis**

**Alternate EJC RIPiTs from HEK293 cells**

RIPiTs were carried out with and without formaldehyde crosslinking as described previously (Singh et al., 2014) with the following modifications. For native RIPiTs, total extracts from four 15-cm plates prepared in hypotonic lysis buffer supplemented with 150mM NaCl-containing were used as input into FLAG-IP. For formaldehyde crosslinked RIPiTs, total extracts were prepared from six 15-cm plates in denaturing lysis buffer supplemented with 150 mM NaCl-containing for input into FLAG-IP. Following IP and washes, RNase I (0.006 U/ml in Isotonic wash buffer (IsoWB)) treatments were performed at 4 °C for 10 min. For the second IP, the following antibodies conjugated to protein-A Dynabeads were used: anti-EIF4A3 (Bethyl A302-980A, 10 µg/RIPiT), anti-CASC3 (Bethyl A302-472A, 2 µg/RIPiT), anti-RNPS1 (HPA044014-100UL, 2 µg/RIPiT). RIPiTs were eluted in clear sample buffer and divided into two parts for RNA and protein analysis as described previously. RIPiTs to enrich EJC footprints upon cycloheximide (CHX) treatment were carried out as above except that cells were incubated with 100 µg/ml CHX for 3 hr prior to harvesting. CHX was included at the same concentration in PBS (for washes before lysis) and cell lysis buffers.

**High-throughput sequencing library preparation**

For RIPiT-Seq, RNA extracted from ~80 % of RIPiT elution was used to generate strand-specific libraries using a custom library preparation method as detailed in (Gangras et al., 2018). For RNA-Seq libraries, 5 µg of total cellular RNA was used as input for ribosomal RNA depletion (Ribozero, Illumina). Purified RNA was then used to generate strand-specific libraries using a custom library preparation method (Gangras et al., 2018). Following PCR amplification, all libraries were quantified using Bioanalyzer (DNA lengths) and Qubit (DNA amounts). Libraries were sequenced on Illumina HiSeq 2500 in the single-end format (50 and 100 nt read lengths).

**Data pre-processing**

*Adapter trimming and PCR duplicate removal*

After demultiplexing, fastq files containing unmapped reads were first trimmed using cutadapt. A 12 nt sequence on read 5' ends consisting of a 5 nt random sequence, 5 nt identifying barcode, and a CC was removed with the random sequence saved for each read for identifying PCR duplicates down the line. Next as much of the 3'-adapter (miR-Cat22) sequence TGGAATTCTCGGGTGCCAAGG was removed from the 3' end as possible. Any reads less than 20 nt in length after trimming were discarded.

*Alignment and removal of multimapping reads*

Following trimming reads were aligned with tophat v2.1.1 (Trapnell et al., 2009) using 12 threads to NCBI GRCh38 with corresponding Bowtie2 index. After alignment reads with a mapping score less than 50 (uniquely mapped) were removed, i.e. all multimapped reads were discarded. Finally all reads mapping to identical regions were compared for their random barcode sequence; if the random sequences matched, such reads were inferred as PCR duplicates and only one such read was kept.

*Removal of stable RNA mapping reads*

Next, reads which came from stable RNA types were counted and removed as follows. All reads were checked for overlap against hg38 annotations for miRNA, rRNA, tRNA, scaRNA, snoRNA, and snRNA using bedtools intersect (Quinlan and Hall, 2010), and any reads overlapping by more than 50 % were removed. Reads aligned to chrM (mitochondrial) were also counted and removed.

**Human reference transcriptome**

The primary reference transcriptome used in all post-alignment analysis was obtained from the UCSC Table Browser. CDS, exon, and intron boundaries were obtained for canonical genes by selecting Track: Gencode v24, Table: knownGene, Filter: knownCanonical (describes the canonical splice variant of a gene).

**Read distribution assignment**

Fractions of reads corresponding to exonic, intronic, intergenic, and canonical EJC and non-canonical EJC regions were then computed. Exonic regions were defined by the canonical hg38 genes, with intronic regions defined as the regions between exons in said genes. Bedtools intersect was used to compare reads against these exon and intron annotations, and reads which overlapped the annotation by more than 50 % were counted. Any reads which did not overlap either the exon or intron annotations sufficiently were counted as intergenic. For classification of reads as canonical versus non-canonical EJC footprints, the canonical region for each library was defined using the meta-exon distribution at exon 3'-ends (Figure 3C and S3J). All reads with their 5'ends falling within the window starting at -24 position till the 25 % max height on the 5' side of the canonical EJC peak were counted as canonical reads. Similarly any read whose 5' end was found anywhere between the start of the exon and 10 bp upstream of that 25 % max point was considered non-canonical.

**k-mer analysis**

Lists of all 6-mers and 3-mers present in reads mapping to exonic regions (as described above) were produced for each RIPiT-Seq sample. The ratio of total 3-mer frequency in RIPiT-Seq samples to RNA-Seq samples was then used to identify 3-mers enriched in alternate EJCs.

**Motif enrichment analysis**

Motif enrichment analysis was performed by first selecting RNA binding proteins of interest - namely SR proteins - from the position weight matrices (PWMs) available on http://rbpdb.ccbr.utoronto.ca/. For all reads mapped to exonic regions, a score was then generated representing the highest possible binding probability for each protein on that read. For visualization the cumulative distribution of these score frequencies was plotted for both pull down and RNA-Seq replicates, with a relatively higher score frequency at a positive score implying greater binding affinity. The p-values between RIPiT-Seq and RNA-Seq replicates were also computed for every score using a negative binomial based model, with significant values primarily in positive score regions implying a binding preference for that protein.

**Differential enrichment analysis**

Differential analysis of exons and transcripts between CASC3 and RNPS1 pull down was conducted with the DESeq2 (Love et al., 2014) package in R. Exons and transcripts with significant differential expression ($p < 0.05$) were selected. All the following analysis was conducted using only the lists of significantly differentially expressed transcripts, unless otherwise noted.

**Estimation of nuclear versus cytoplasmic levels**

Nuclear and cytoplasmic RNA levels were estimated by first obtaining nuclear and cytoplasmic reads from (Neve et al., 2016). Reads were aligned and mapped to our exonic annotation as described above, and a ratio of nuclear to cytoplasmic reads was then calculated for all transcripts.

**Comparison to genes with detained introns**

A list of detained and non-detained introns was obtained from (Boutz et al., 2015). To identify detained intron containing and lacking genes in our RNA-Seq data, we carried out analysis using a DESeq2-based pipeline as described by Boutz *et al*. Briefly, using two RNA-Seq replicates we first created artificial data sets containing the same numbers of total reads per replicate, but with counts originating from introns in a given gene spread evenly amongst those introns in the artificial set. By comparing the intron distributions of these artificial replicates to the experimental replicates using DESeq we were able to produce lists of detained introns (introns with significantly ($p < 0.05$) higher expression levels compared to the artificially spread data) and non-detained introns (introns with non-significant ($p > 0.1$) expression levels compared to the artificially spread data). The transcripts containing introns found to be detained/non-detained in both our analysis and the analysis done by Boutz *et al*. make up the stringent lists, which were used for analysis in Figure 4D. Of the 693 canonical genes containing detained introns reported by Boutz et al. we found 555 in our own analysis (80 %) and of the 5294 canonical genes lacking detained introns we found 812 (15 %).

**Features compared between two alternate EJC sets**

mRNA half life data was taken from (Tani et al., 2012) with no further processing on our part. Translation efficiency data was similarly obtained from (Kiss et al., 2017).

**Gene ontology analysis**

DAVID gene ontology tool (Huang et al., 2009) was used to compare the set of genes (canonical Ensembl transcript IDs) predicted by DESeq2 analysis to be significantly enriched in CASC3 or RNPS1 EJCs against a background list containing only those human genes that were reliably detected by DESeq2 (all genes for which DESeq2 calculated adjusted p-values). Only non-redundant categories with lowest p-value (with Benjamini-Hochberg correction) are reported.